\documentclass[11pt]{article}
\usepackage{epsfig} 
\newcommand{\sect}[1]{ \section{#1} \setcounter{equation}{0} }

\newcommand{\Dslash}{D \! \! \! \! /}

\newcommand{\half}{\mbox{\small{$\frac{1}{2}$}}}

\newcommand{\MSbar}{\overline{\mbox{MS}}} 
 
\newcommand{\Nf}{N_{\!f}}
\newcommand{\NF}{N_{\!F}}
\newcommand{\NA}{N_{\!A}}

\begin{document}
\title{Alternative refined Gribov-Zwanziger Lagrangian}
\author{J.A. Gracey, \\ Theoretical Physics Division, \\ 
Department of Mathematical Sciences, \\ University of Liverpool, \\ P.O. Box 
147, \\ Liverpool, \\ L69 3BX, \\ United Kingdom.} 
\date{}
\maketitle 

\vspace{5cm} 
\noindent 
{\bf Abstract.} We consider the implications of the condensation of a general
local BRST invariant dimension two operator built out of the localizing ghost 
fields of the Gribov-Zwanziger Lagrangian which is a localized Lagrangian
incorporating the Gribov problem in the Landau gauge. For different colour 
tensor projections of the general operator, the properties of a frozen gluon 
propagator and unenhanced Faddeev-Popov ghost propagator, which are observed in 
lattice computations, can be reproduced. The alternative possibilities are 
distinguished by the infrared structure of the propagators of the spin-$1$
fields, other than those of the gluon and Faddeev-Popov ghost, for which there 
is no numerical simulation data to compare with yet.  

\vspace{-16.7cm}
\hspace{13.5cm}
{\bf LTH 884}
 
\newpage

\sect{Introduction.}

Yang-Mills theories play an important role in our understanding of the 
fundamental particles of nature. For instance, the strong force is believed to
be reliably described by an $SU(3)$ non-abelian gauge theory whose fundamental
particles are quarks and gluons. In nature these particles cannot be isolated 
as separate entities. Instead at high energies they are asymptotically free,
\cite{1,2}, but at low energies they are present only within bound states of 
hadrons. Understanding the specific dynamics of the mechanism which confines 
quarks and gluons and prevents them from being observed in nature is currently 
a major topic of interest. Ordinarily in a quantum field theory the behaviour 
of a particle's propagator plays a key role in its interpretation as an 
observed quanta. For instance, in Quantum Electrodynamics (QED) the electron 
propagator has a simple pole at real positive value of its squared momentum 
which corresponds to its physical mass when all quantum corrections have been 
computed to all orders in perturbation theory. The form of this fundamental 
propagator is therefore associated with a real observed particle. By contrast
in Quantum Chromodynamics (QCD), which is the non-abelian generalization for 
the strong force, the propagators of the quark and gluon fields derived from 
the canonical Lagrangian emerge as being of the same fundamental form but in 
the case of the gluon it is massless. Clearly this cannot be correct since that
would imply that the gluon should exist as a free massless observed particle.
However, this analysis derives from a gauge fixed Lagrangian with a non-abelian
symmetry. It transpires that contrary to what occurs in an abelian theory there
are problems in fixing the gauge uniquely globally in Yang-Mills theories. This
was first pointed out in detail by Gribov in his seminal work, \cite{3}. In 
essence one can only fix a gauge uniquely locally in a non-abelian gauge theory
but not globally. It turns out that one can construct different gauge 
configurations satisfying the same gauge condition in the non-abelian case. 
Therefore, there is an overcounting in the definition of the path integral and 
the region of integration in configuration space has to be restricted to a 
specific subspace. This is known as the first Gribov region and it contains the
origin, \cite{3}. Whilst this overcomes a significant amount of the copy issue 
the gauge is still not uniquely fixed. Only inside the fundamental modular 
region which is within this horizon is there a unique globally fixed gauge. 

In analysing this problem for the Landau gauge Gribov, \cite{3}, managed to 
determine several interesting features. First, the path integral could be 
modified to incorporate the restriction to the Gribov region. The boundary is 
defined by the no-pole condition where the Faddeev-Popov operator vanishes. 
This means that the inverse is finite within the horizon of the Gribov region. 
Using a semi-classical analysis Gribov demonstrated that the path integral 
cut-off modified the gluon propagator. In particular the fundamental behaviour 
was replaced by a propagator which vanished at zero momentum, known as gluon 
suppression, but the simple massless pole property accepted at high energy was
retained. Underlying this was a new mass parameter called the Gribov mass and 
denoted by $\gamma$, \cite{3}. It is not an independent quantity but is a 
function of the coupling constant, $g$, and defined via the horizon condition 
cutting off the path integral leading to a gap equation. Only when the Gribov
mass actually satisfies the gap equation can the theory be regarded as a gauge 
theory. This gap equation was responsible for another novel feature which is 
that the Faddeev-Popov ghost propagator behaved as a dipole in the infrared 
which is known as ghost enhancement. This latter property was subsequently 
reformulated in terms of the Kugo-Ojima confinement criterion, \cite{4,5}, 
which was established by a BRST analysis of the Landau gauge. However, one of 
the main consequences is that the analysis showed that the gluon propagator 
ceased to be fundamental when the copy problem was taken into consideration and
the hope was that the Gribov problem was fundamental to the problem of 
confinement, \cite{3}.

One of the main difficulties in extending the semi-classical analysis of
\cite{3} was that the no pole condition introduces a non-local operator into 
the Lagrangian. This is an obstacle to performing canonical field theory 
calculations which only proceed when a Lagrangian is at least local and
renormalizable. However, in a series of articles, 
\cite{6,7,8,9,10,11,12,13,14,15,16}, Zwanziger overcame this by managing to not
only rewrite the Gribov Lagrangian in a local form but also in a way that it 
was renormalizable, \cite{13,17,18}. The localization proceeded by the 
introduction of a set of additional ghost fields, $\{ \phi^{ab}_\mu, 
\bar{\phi}^{ab}_\mu, \omega^{ab}_\mu, \bar{\omega}^{ab}_\mu \}$. The first pair
are bosonic whilst the second set have Fermi statistics and are necessary to 
ensure the theory remains asymptotically free at high energy, for instance. The
renormalization group functions are not affected by these additional fields 
whose effect is only apparent in the infrared region, \cite{13,17,18}. 
Moreover, no extra renormalization constants are required since the anomalous
dimensions of the localizing ghosts are the same as that of the Faddeev-Popov 
ghost and $\gamma$ is rendered finite merely by a simple combination of the 
gluon and ghost renormalization constants. With these additions to the 
Lagrangian the original suppressed gluon propagator is retained. However, one 
can study the loop corrections using the localized Gribov version which is 
known as the Gribov-Zwanziger Lagrangian. For instance, the one loop gap 
equation can be computed, \cite{8,13}, and reproduces that of Gribov, \cite{3}.
This was extended to two loops in the $\MSbar$ scheme in \cite{19} with the one
loop propagator corrections also deduced in \cite{20,21}. As noted in 
\cite{11,12} by Zwanziger the gluon remains suppressed when quantum corrections
are included. Also the Kugo-Ojima criterion was shown to be satisfied at two 
loops, \cite{19,20}. Subsequently this criterion was revisited but in the 
context of the Gribov-Zwanziger Lagrangian, \cite{22}, where the additional 
localizing fields have to be included in repeating the original BRST analysis 
of \cite{4,5}. It was shown that the Kugo-Ojima condition was valid in the 
Gribov-Zwanziger Lagrangian.

Having summarized the theoretical analysis of accommodating the gauge copy 
problem in the Landau gauge in QCD the next issue in this area relates to 
lattice data collected over the last few years on the gluon and Faddeev-Popov 
ghost propagators. Given the advances in computing power and algorithms to 
carry out the gauge fixing numerically on larger lattices, 
\cite{23,24,25,26,27,28,29,30,31}, there is now a reasonable amount of data on 
both propagators at low momenta. More specifically, the zero momentum behaviour
appears to show clearly that contrary to the Gribov scenario the gluon 
propagator freezes to a {\em non-zero} value and the Faddeev-Popov ghost does 
not enhance. Instead its infrared form does not significantly deviate from that
of its ultraviolet fundamental massless form. Indeed there is also similar 
evidence from Schwinger-Dyson computations. See, for example, \cite{31,32,33}. 
Thus there are two distinct cases. One is the gluon suppressed but ghost 
enhanced propagators known as the scaling or conformal solution. The other is 
the non-zero frozen gluon propagator with the unenhanced ghost. Such a gluon
propagator was originally derived in \cite{33} in a Schwinger-Dyson analysis
and referred to as the massive solution. More recently, \cite{34}, the term 
decoupling solution has been used. As both labels are recognised we will use 
them synonymously in the discussion. Therefore, there has been a debate as to 
how the numerical data fit in with the Gribov scenario, \cite{3}, and whether 
the absence of the Kugo-Ojima confinement was significant. One point of view 
was that satisfying the criterion or not was an additional condition on the 
gauge fixing procedure, \cite{35}, effective in the infrared. However, given 
the properties and elegance of the Gribov-Zwanziger Lagrangian it seemed 
appropriate to try and extract a gluon propagator with non-zero freezing and a 
non-enhanced Faddeev-Popov ghost in that approach. This was achieved in the 
series of articles, \cite{36,37,38}. In essence the Lagrangian was refined to 
include a local BRST invariant dimension two operator built from the localizing
ghost fields. Ordinarily such a term would just introduce masses for these 
fields only. However, the complicated way the localization arises means that in
constructing the propagators from the quadratic sector the extra mass term 
affects the gluon propagator too. Specifically the extra mass parameter appears
in the gluon propagator in such as way that there is non-zero freezing at zero 
momentum. Equally the extra mass excludes the enhancement of the Faddeev-Popov 
ghost and therefore the numerical data and Schwinger-Dyson analyses can be 
modelled by what was termed a refinement of the original Gribov-Zwanziger 
Lagrangian. The justification for the inclusion of this additional mass 
operator is to consider the condensation of the operator via a dynamical 
mechanism. The analysis for this is achieved by the local composite operator 
(LCO) formalism, \cite{39,40}, which was initially used in QCD to study the 
condensation of a gluon mass operator, \cite{41,42}. 

Whilst the computation of \cite{36,37,38} clearly accommodates the current 
picture of the gluon and ghost propagators it transpires that the most general 
dimension two BRST operator was not considered. Given that the localizing
fields carry more than one colour adjoint index there are six different colour 
projections. Therefore, it is the purpose of this article to perform a 
comprehensive analysis of the various different colour channels. It will turn 
out that there are other operators whose condensation will lead to a non-zero 
freezing of the gluon propagator and an unenhanced Faddeev-Popov ghost 
propagator. Therefore, the refined Gribov-Zwanziger Lagrangian of 
\cite{36,37,38} is not the {\em unique} model of the lattice data which is why 
we will refer to the explanations here as an alternative. Moreover, in our 
results we will show that the conformal or scaling behaviour of the original or
pure Gribov-Zwanziger analysis can persist even with the extra mass operator. 
Although the alternative we focus on here, which will be referred to as the 
${\cal R}$ channel for reasons which will be clear later, is equivalent to the 
earlier explanation, which will be called the ${\cal Q}$ channel, the full 
infrared behaviour of both are not equivalent. For instance, the behaviour of 
the Bose ghost propagator is different in each channel. The study of this 
particular propagator has been of interest recently in the pure 
Gribov-Zwanziger Lagrangian, \cite{43,21}. It has been shown that there is 
enhancement in certain colour channels. This was observed at the one loop 
level, \cite{21}, but proved to be an all orders feature in \cite{43}. Briefly,
the enhanced Bose fields are the Goldstone bosons associated with the 
spontaneous breaking of the BRST symmetry in a theory where the fields are 
constrained by the horizon condition, \cite{43}. In principle one could resolve
which of the ${\cal Q}$ or ${\cal R}$ cases was correct when the BRST invariant
operator is included if there was numerical data for this propagator. However, 
these localizing ghost fields only exist within the Gribov-Zwanziger Lagrangian 
itself and not within the lattice formulation of QCD. Therefore, it is not 
clear whether one could construct the equivalent correlation function in the 
lattice language. Whilst we will carry out the one loop analysis for the 
${\cal R}$ channel we will also give a general indication of the properties of 
the additional colour channel possibilities in order to have as a complete 
picture as possible. This would become important if additional numerical data 
became available which rules out either of the ${\cal Q}$ or ${\cal R}$ channel
explanations. Interestingly in certain scenarios the colour tensor structure 
which was observed in \cite{43,21}, but for the enhanced Bose localizing 
propagator at one loop, actually arises in the dominant term in the zero 
momentum limit in the original propagator prior to any loop analysis.
 
The article is organised as follows. The key properties of the pure
Gribov-Zwanziger Lagrangian are reviewed in the next section. The most general 
local BRST invariant dimension two operator built from the localizing ghosts is
introduced in section three and the propagators are constructed for each of the
six potential single additional mass parameters. The justification for the 
inclusion of such additional masses is provided in the subsequent section where
the focus is on the ${\cal R}$ channel. There the one loop effective potential 
for the operator is constructed as well as the one loop $\MSbar$ corrections to
all the propagators. These are required to study the effect of the gap equation
satisfied by $\gamma$ has on the propagator corrections in order to see, for 
example, whether there is any Bose ghost enhancement. Our conclusions are given
in section five. There are two appendices. The first gives the full propagators
for the inclusion of all six colour channel projections of the operator for the
case of $SU(3)$ whilst the second records the propagators for the ${\cal W}$ 
and ${\cal S}$ channels separately for $SU(N_c)$. Whilst the expressions 
recorded in both appendices for the propagators are cumbersome they are 
significantly compact compared with those for the general colour group case.

\sect{Background.}

We begin by briefly recalling the essential features of the Gribov-Zwanziger
Lagrangian and the construction of the propagators. Throughout the article we
work in the Landau gauge. The basic Lagrangian is, \cite{3},  
\begin{equation}
L^{\mbox{\footnotesize{Grib}}} ~=~ L^{\mbox{\footnotesize{QCD}}} ~+~
\frac{C_A \gamma^4}{2} A^{a\,\mu} \frac{1}{\partial^\nu D_\nu} A^a_\mu ~-~
\frac{d \NA \gamma^4}{2g^2}
\end{equation}
where the original gauge fixed Lagrangian is
\begin{equation} 
L^{\mbox{\footnotesize{QCD}}} ~=~ -~ \frac{1}{4} G_{\mu\nu}^a 
G^{a \, \mu\nu} ~-~ \frac{1}{2\alpha} (\partial^\mu A^a_\mu)^2 ~-~ 
\bar{c}^a \partial^\mu D_\mu c^a ~+~ i \bar{\psi}^{iI} \Dslash \psi^{iI} 
\label{lagqcd}
\end{equation} 
and $\alpha$ is the covariant gauge fixing parameter. Although we have included
$\alpha$ we will focus purely on the Landau gauge, corresponding to
$\alpha$~$=$~$0$, but it is retained here since it is crucial in obtaining the
propagators. The parameter $\gamma$ is the Gribov mass which is not an 
independent quantity but is the mass scale which is implicit in the horizon 
condition, \cite{3},  
\begin{equation}
\left\langle A^a_\mu(x) \frac{1}{\partial^\nu D_\nu} A^{a\,\mu}(x)
\right\rangle ~=~ \frac{d N_A}{C_A g^2}
\label{hordef}
\end{equation}
where $\NA$ is the dimension of the adjoint representation. The presence of the
horizon condition alters the infrared behaviour of the gluon and Faddeev-Popov
propagators in the original Gribov scenario, \cite{3}. As has already been well
documented the non-local term renders practical computations impossible. 
However, Zwanziger's series of articles on how one can localize the
non-locality produces a renormalizable local Lagrangian, 
\cite{7,8,11,12,13,14,44}, which is 
\begin{eqnarray}
L^{\mbox{\footnotesize{GZ}}} &=& L^{\mbox{\footnotesize{QCD}}} ~+~ 
\frac{1}{2} \rho^{ab \, \mu} \partial^\nu \left( D_\nu \rho_\mu 
\right)^{ab} ~+~ \frac{i}{2} \rho^{ab \, \mu} \partial^\nu 
\left( D_\nu \xi_\mu \right)^{ab} ~-~ \frac{i}{2} \xi^{ab \, \mu} 
\partial^\nu \left( D_\nu \rho_\mu \right)^{ab} \nonumber \\
&& +~ \frac{1}{2} \xi^{ab \, \mu} \partial^\nu \left( D_\nu \xi_\mu 
\right)^{ab} ~-~ \bar{\omega}^{ab \, \mu} \partial^\nu \left( D_\nu \omega_\mu 
\right)^{ab} ~-~ \frac{1}{\sqrt{2}} g f^{abc} \partial^\nu 
\bar{\omega}^{ae}_\mu \left( D_\nu c \right)^b \rho^{ec \, \mu} \nonumber \\
&& -~ \frac{i}{\sqrt{2}} g f^{abc} \partial^\nu \bar{\omega}^{ae}_\mu 
\left( D_\nu c \right)^b \xi^{ec \, \mu} ~-~ i \gamma^2 f^{abc} A^{a \, \mu} 
\xi^{bc}_\mu ~-~ \frac{d \NA \gamma^4}{2g^2} ~. 
\label{laggz}
\end{eqnarray} 
There are additional fields to the gluon, $A^a_\mu$, Faddeev-Popov ghost,
$c^a$, and massless quark, $\psi^{iI}$, which are two Bose ghosts,
$\xi^{ab}_\mu$ and $\rho^{ab}_\mu$, and the Grassmann ghosts $\omega^{ab}_\mu$ 
and $\bar{\omega}^{ab}_\mu$. The indices range over the values
$1$~$\leq$~$a$~$\leq$~$\NA$, $1$~$\leq$~$I$~$\leq$~$\NF$ and
$1$~$\leq$~$i$~$\leq$~$\Nf$ where $\NF$ is the dimension of the fundamental 
representation and $\Nf$ is the number of massless quarks. In earlier localized
versions of the Gribov Lagrangian the complexified Bose ghosts $\phi^{ab}_\mu$ 
and $\bar{\phi}^{ab}_\mu$ were used. Given recent developments, \cite{43,44}, 
in terms of the propagator structure of the real versions of the Bose ghosts we
will use this formulation but note that the relation between the two versions 
are trivial and given by  
\begin{equation}
\phi^{ab}_\mu ~=~ \frac{1}{\sqrt{2}} \left( \rho^{ab}_\mu ~+~ i \xi^{ab}_\mu 
\right) ~~,~~ 
\bar{\phi}^{ab}_\mu ~=~ \frac{1}{\sqrt{2}} \left( \rho^{ab}_\mu ~-~ 
i \xi^{ab}_\mu \right) ~.
\label{realcomp}
\end{equation} 
However, the renormalizability proofs of the Lagrangian, \cite{13,17,18}, were
performed for the complex Bose ghost field and used the BRST symmetry of the 
localized Lagrangian. Since we will be considering a BRST dimension two 
operator we note the BRST symmetry transformations for (\ref{laggz}) given
(\ref{realcomp}) are 
\begin{eqnarray}
\delta A^a_\mu &=& -~ \left( D_\mu c \right)^a ~~~,~~~ 
\delta c^a ~=~ \frac{1}{2} f^{abc} c^b c^c ~~~,~~~
\delta \bar{c}^a ~=~ b^a ~~~,~~~ \delta b^a ~=~ 0 \nonumber \\
\delta \phi^{ab}_\mu &=& \omega^{ab}_\mu ~~~,~~~
\delta \omega^{ab}_\mu ~=~ 0 ~~~,~~~
\delta \bar{\phi}^{ab}_\mu ~=~ 0 ~~~,~~~
\delta \bar{\omega}^{ab}_\mu ~=~ \bar{\phi}^{ab}_\mu
\end{eqnarray}
where $b^a$ is the Nakanishi-Lautrup auxiliary field. Therefore, it is trivial
to observe that the colour non-singlet operator
\begin{equation}
{\cal O}^{abcd} ~=~ \bar{\phi}^{ab} \phi^{cd} ~-~ \bar{\omega}^{ab} \omega^{cd} 
\label{brstopc}
\end{equation}
is BRST invariant. Rewriting in terms of the real Bose ghosts the operator is
equivalent to  
\begin{equation}
{\cal O}^{abcd} ~=~ \frac{1}{2} \left[ \rho^{ab} \rho^{cd} ~+~ 
i \xi^{ab} \rho^{cd} ~-~ i \rho^{ab} \xi^{cd} ~+~ \xi^{ab} \xi^{cd} \right] ~-~
\bar{\omega}^{ab} \omega^{cd} ~. 
\label{brstop}
\end{equation}
In \cite{36,37,38} it was the condensation of the operator $\delta^{ac} 
\delta^{bd} {\cal O}^{abcd}$ which was investigated using the local composite 
operator formalism in order to see what effect it had on the infrared structure
of the propagators. 

In this localized Lagrangian the horizon condition, (\ref{hordef}), is replaced
by the vacuum expectation value of local fields, \cite{43,44}, 
\begin{equation}
f^{abc} \left\langle A^{a\,\mu} (x) \xi^{bc}_\mu(x) \right\rangle ~=~ 
\frac{i d\NA \gamma^2}{g^2} 
\label{hordefloc}
\end{equation}
which determines $\gamma$ to two loops in the $\MSbar$ scheme, \cite{19}, by 
the solution of 
\begin{eqnarray}
1 &=& C_A \left[ \frac{5}{8} - \frac{3}{8} \ln \left(
\frac{C_A\gamma^4}{\mu^4} \right) \right] a \nonumber \\
&& +~ \left[ C_A^2 \left( \frac{3893}{1536} - \frac{22275}{4096} s_2
+ \frac{29}{128} \zeta(2)
- \frac{65}{48} \ln \left( \frac{C_A\gamma^4}{\mu^4} \right)
+ \frac{35}{128} \left( \ln \left( \frac{C_A\gamma^4}{\mu^4} \right)
\right)^2 \right. \right. \nonumber \\
&& \left. \left. ~~~~~~~~~~~~+~ \frac{411}{1024} \sqrt{5} \zeta(2)
- \frac{1317\pi^2}{4096} \right) \right. \nonumber \\
&& \left. ~~~~~+~ C_A T_F \Nf \left( \frac{\pi^2}{8} - \frac{25}{24} - \zeta(2)
+ \frac{7}{12} \ln \left( \frac{C_A\gamma^4}{\mu^4} \right)
- \frac{1}{8} \left( \ln \left( \frac{C_A\gamma^4}{\mu^4} \right) \right)^2
\right) \right] a^2 \nonumber \\
&& +~ O(a^3)
\label{gap2}
\end{eqnarray}
where $\zeta(z)$ is the Riemann zeta function, $s_2$~$=$~$(2\sqrt{3}/9) 
\mbox{Cl}_2(2\pi/3)$ and $\mbox{Cl}_2(x)$ is the Clausen function and 
$a$~$=$~$g^2/(16\pi^2)$. However, it was demonstrated recently that the horizon
condition could be replaced by a vacuum expectation value of an operator 
involving only the Bose ghost, albeit one involving an infinite number of 
terms, \cite{45}. Specifically, 
\begin{eqnarray}
&& f_4^{abcd} \left\langle \partial^\nu
\xi^{ab\,\mu} \left[ \partial_\nu \xi^{cd}_\mu ~-~ \frac{i g}{C_A \gamma^2}
f_4^{cfrs} \left( \partial^\sigma \partial_\sigma \xi^{rs}_\nu \right)
\xi^{fd}_\mu \right. \right. \nonumber \\
&& \left. \left. ~~~~~~~~~~~~~~~~~~~-~ \frac{g^2}{C_A^2 \gamma^4} f_4^{cfrs}
f_4^{rqmn} \partial^\sigma \left[ \left( \partial^\rho \partial_\rho
\xi^{mn}_\sigma \right) \xi^{qs}_\nu \right] \xi^{fd}_\mu \right. \right.
\nonumber \\
&& \left. \left. ~~~~~~~~~~~~~~~~~~~+~ O(g^3) \right] \right\rangle
~=~ \frac{d C_A \NA \gamma^4}{g^2}
\end{eqnarray}
where $f_4^{abcd}$~$\equiv$~$f^{abp} f^{cdp}$. This reproduces (\ref{gap2}) to
two loops and is constructed by application of the equation of motion
\begin{equation}
A^a_\mu ~=~ -~ \frac{i}{C_A \gamma^2} f^{abc} \left( \partial^\nu D_\nu \xi_\mu
\right)^{bc} 
\label{eom}
\end{equation}
to (\ref{hordefloc}) via the intermediate vacuum expectation value
\begin{equation}
f^{abp} f^{cdp} \left\langle \xi^{ab\,\mu} (x) \left( \partial^\nu D_\nu
\xi_\mu \right)^{cd} (x) \right\rangle ~=~ -~ \frac{d C_A \NA \gamma^4}{g^2} 
\label{gapdefalt}
\end{equation}
where $d$ is the spacetime dimension.

One of the reasons for reviewing the various formulations of the gap equation 
is to motivate our alternative mechanism for modelling the massive or 
decoupling solution observed on the lattice. First, we recall that the Landau 
gauge propagators for the spin-$1$ fields of (\ref{laggz}) are 
\begin{eqnarray}
\langle A^a_\mu(p) A^b_\nu(-p) \rangle &=& -~ 
\frac{\delta^{ab}p^2}{[(p^2)^2+C_A\gamma^4]} P_{\mu\nu}(p) \nonumber \\
\langle A^a_\mu(p) \xi^{bc}_\nu(-p) \rangle &=& 
\frac{i f^{abc}\gamma^2}{[(p^2)^2+C_A\gamma^4]} P_{\mu\nu}(p) 
\nonumber \\
\langle A^a_\mu(p) \rho^{bc}_\nu(-p) \rangle &=& 0 \nonumber \\ 
\langle \xi^{ab}_\mu(p) \xi^{cd}_\nu(-p) \rangle &=& -~ 
\frac{\delta^{ac}\delta^{bd}}{p^2}\eta_{\mu\nu} ~+~
\frac{f^{abe}f^{cde}\gamma^4}{p^2[(p^2)^2+C_A\gamma^4]} P_{\mu\nu}(p) 
\nonumber \\ 
\langle \xi^{ab}_\mu(p) \rho^{cd}_\nu(-p) \rangle &=& 0 \nonumber \\ 
\langle \rho^{ab}_\mu(p) \rho^{cd}_\nu(-p) \rangle &=& 
\langle \omega^{ab}_\mu(p) \bar{\omega}^{cd}_\nu(-p) \rangle ~=~ -~ 
\frac{\delta^{ac}\delta^{bd}}{p^2} \eta_{\mu\nu} 
\label{prop}
\end{eqnarray} 
where 
\begin{equation}
P_{\mu\nu}(p) ~=~ \eta_{\mu\nu} ~-~ \frac{p_\mu p_\nu}{p^2} ~~~,~~~
L_{\mu\nu}(p) ~=~ \frac{p_\mu p_\nu}{p^2} 
\end{equation}
are the usual transverse and longitudinal projectors. Therefore to reproduce 
the leading order contribution to the gap equation, and hence obtain Gribov's 
original expression, one simply integrates the mixed $A^a_\mu$-$\xi^{ab}_\mu$ 
propagator of (\ref{prop}). Equally the alternative formulation of the gap
equation involving only the Bose ghost can be justified at leading order by
noting that the second term of the $\xi^{ab}_\mu$ propagator needs the massless
pole to be absent. This is achieved by the inclusion of the wave operator,
$\partial^\sigma \partial_\sigma$, at leading order, \cite{3}, and eventually 
correctly with the Faddeev-Popov operator in keeping with the ethos of 
Zwanziger's extension of Gribov's semi-classical horizon condition, \cite{7}. 
Given this, one can obtain an idea of which of the other operators may condense
and be dominant or relevant in the infrared at leading order by considering 
other propagators of (\ref{prop}). Clearly integrating the gluon propagator 
suggests the condensation of the gauge variant operator $\half A^a_\mu 
A^{a\,\mu}$ which has been widely treated. However, if one considers the BRST 
invariant operator, (\ref{brstop}), then integrating the relevant Bose and 
Grassmann propagators of (\ref{prop}) we find the colour dependence 
\begin{equation}
\langle {\cal O}^{abcd} \rangle ~ \propto ~ f^{abe} f^{cde} ~.
\label{condop}
\end{equation}
Clearly the other colour channel of the relevant propagators, $\delta^{ac}
\delta^{bd}$, gives zero upon integration in dimensional regularization. Whilst
we are not saying that there could be no condensation in this colour channel, 
or any others not present here, we are suggesting that the more natural colour
channel to condense is that of (\ref{condop}). It is natural in the sense that
it does not differ too much from the original localized Lagrangian structure.
Moreover, it would appear to be the dominant expectation value at leading
order. Given this simple observation we will examine the consequences for this 
taking this colour channel choice in (\ref{brstop}) and discuss alternative
cases as well in the next section.

\sect{Propagators.}

In this section we consider the construction of the spin-$1$ propagators for 
(\ref{laggz}) when there are extra mass terms eminating from a BRST invariant 
mass. As ${\cal O}^{abcd}$ is BRST invariant then in order to have a colour 
scalar it needs to be contracted with a rank $4$ colour tensor. To be as
flexible as possible at the outset we take the general operator
\begin{equation}
{\cal O} ~=~ \left[ \mu_{{\cal Q}}^2 \delta^{ac} \delta^{bd} ~+~ 
\mu_{{\cal W}}^2 f^{ace} f^{bde} ~+~ \frac{\mu_{{\cal R}}^2}{C_A} 
f^{abe} f^{cde} ~+~ \mu_{{\cal S}}^2 d_A^{abcd} ~+~ 
\frac{\mu_{{\cal P}}^2}{\NA} \delta^{ab} \delta^{cd} ~+~ \mu_{{\cal T}}^2 
\delta^{ad} \delta^{bc} \right] {\cal O}^{abcd}
\label{brstgen}
\end{equation}
where $\mu_i^2$ are the various mass parameters. The labelling is chosen in
order to track the location of the various colour structures within the final 
propagators. The inclusion of $C_A$ and $\NA$ in several terms is in order to
have a degree of uniformity in the various different final propagator forms and
to make relative comparisons easy to follow as will be evident later. In 
addition to be complete we include an additional dimension two operator which 
is
\begin{equation}
{\cal O}_{A_\mu^2} ~=~ \frac{\mu_{{\cal X}}^2}{2} A^a_\mu A^{a\,\mu} ~.
\label{opaa}
\end{equation}
In a general linear covariant gauge for (\ref{lagqcd}) the operator
$\half A^a_\mu A^{a\,\mu}$~$-$~$\alpha \bar{c}^a c^a$ is BRST invariant and was
introduced as a potential gluon mass operator by Curci and Ferrari in 
\cite{46}. So in the Landau gauge (\ref{opaa}) represents a natural additional
operator in the context of studying the BRST operator for the localizing ghost 
sector. So it seems appropriate to include it in our current construction to 
ascertain its effect. The subscripts, ${\cal X}$, ${\cal Q}$, ${\cal W}$, 
${\cal R}$, ${\cal S}$, ${\cal P}$ and ${\cal T}$, derive from notation used in
earlier papers, \cite{20,21}. There the colour structure of the one loop 
corrections to the $2$-point functions was examined and the inversion of the 
appropriate matrix of $2$-point functions was performed to obtain the one loop 
propagator corrections. This involved understanding the multiplication of the 
colour tensors associated with each amplitude. As the colour structure for the 
BRST invariant mass term is now of the most general form, we will use the same 
approach as \cite{20,21} to construct the propagators and recall the essentials
for this.

First, for the spin-$1$ sector we focus on the fields which mix through the
term involving $\gamma^2$ in (\ref{laggz}). Therefore, we take our basis of
fields to be $\{A^a_\mu,\xi^{ab}_\mu,\rho^{ab}_\mu\}$ in that order. Although 
there is no mixing with $\rho^{ab}_\mu$ we have included it here since there is
potentially an effect in the longitudinal sector. Next we define the matrix of 
quadratic terms in the momentum space version of the Lagrangian as, \cite{20}, 
\begin{equation}
\Lambda^{\{ab|cd\}} ~=~ \left(
\begin{array}{ccc}
{\cal X} \delta^{ac} & {\cal U} f^{acd} & 0 \\
{\cal U} f^{cab} & {\cal Q}^{abcd}_\xi & 0 \\
0 & 0 & {\cal Q}^{abcd}_\rho \\
\end{array}
\right) 
\end{equation} 
for the transverse sector and 
\begin{equation}
\Lambda^{L\,\{ab|cd\}} ~=~ \left(
\begin{array}{ccc}
{\cal X}^L \delta^{ac} & {\cal U}^L f^{acd} & {\cal V}^L f^{acd} \\
{\cal U}^L f^{cab} & {\cal Q}^{L\, abcd}_\xi & 0 \\
{\cal V}^L f^{cab} & 0 & {\cal Q}^{L\, abcd}_\rho \\
\end{array}
\right) 
\end{equation} 
for the longitudinal sector where we will use the superscript $L$ throughout to
differentiate it from the transverse sector. The respective Lorentz projectors
are passive and omitted, \cite{20,21}. The colour decompositions are taken to 
be 
\begin{eqnarray}
{\cal Q}^{abcd}_\xi &=& {\cal Q}_\xi \delta^{ac} \delta^{bd} ~+~ {\cal W}_\xi
f^{ace} f^{bde} ~+~ {\cal R}_\xi f^{abe} f^{cde} ~+~ 
{\cal S}_\xi d_A^{abcd} ~+~ {\cal P}_\xi \delta^{ab} \delta^{cd} ~+~ 
{\cal T}_\xi \delta^{ad} \delta^{bc} \nonumber \\ 
{\cal Q}^{abcd}_\rho &=& {\cal Q}_\rho \delta^{ac} \delta^{bd} ~+~ 
{\cal W}_\rho f^{ace} f^{bde} ~+~ {\cal R}_\rho f^{abe} f^{cde} ~+~ 
{\cal S}_\rho d_A^{abcd} ~+~ {\cal P}_\rho \delta^{ab} \delta^{cd} ~+~ 
{\cal T}_\rho \delta^{ad} \delta^{bc} \nonumber \\ 
\end{eqnarray}
which is the origin of our earlier notation and 
\begin{eqnarray}
{\cal Q}^{L\, abcd}_\xi &=& {\cal Q}^L_\xi \delta^{ac} \delta^{bd} ~+~ 
{\cal W}^L_\xi f^{ace} f^{bde} ~+~ {\cal R}^L_\xi f^{abe} f^{cde} ~+~ 
{\cal S}^L_\xi d_A^{abcd} ~+~ {\cal P}^L_\xi \delta^{ab} \delta^{cd} ~+~ 
{\cal T}^L_\xi \delta^{ad} \delta^{bc} \nonumber \\ 
{\cal Q}^{L\, abcd}_\rho &=& {\cal Q}^L_\rho \delta^{ac} \delta^{bd} ~+~ 
{\cal W}^L_\rho f^{ace} f^{bde} ~+~ {\cal R}^L_\rho f^{abe} f^{cde} ~+~ 
{\cal S}^L_\rho d_A^{abcd} ~+~ {\cal P}^L_\rho \delta^{ab} \delta^{cd} ~+~ 
{\cal T}^L_\rho \delta^{ad} \delta^{bc} \nonumber \\
\end{eqnarray}
with 
\begin{equation}
d_A^{abcd} ~=~ \frac{1}{6} \mbox{Tr} \left( T_A^a T_A^{(b} T_A^c T_A^{d)}
\right)
\end{equation}
being the rank $4$ totally symmetric tensor in the adjoint representation,
\cite{47}. The propagators are then obtained by inverting the quadratic part
of the momentum space Lagrangian. However, this is more involved than usual due
to the colour structure and is formally given by the corresponding colour
structures for the transverse sector,
\begin{equation}
\Pi^{\{cd|pq\}} ~=~ \left(
\begin{array}{ccc}
{\cal A} \delta^{cp} & {\cal B} f^{cpq} & 0 \\
{\cal B} f^{pcd} & {\cal D}^{cdpq}_\xi & 0 \\
0 & 0 & {\cal D}^{cdpq}_\rho \\
\end{array}
\right) 
\end{equation} 
and 
\begin{equation}
\Pi^{L\,\{cd|pq\}} ~=~ \left(
\begin{array}{ccc}
{\cal A}^L \delta^{cp} & {\cal B}^L f^{cpq} & {\cal C}^L f^{cpq} \\
{\cal B}^L f^{pcd} & {\cal D}^{L\, cdpq}_\xi & {\cal E}^{L\, cdpq}_\xi \\
{\cal C}^L f^{pcd} & {\cal E}^{L\, cdpq}_\xi & {\cal E}^{L\, cdpq}_\rho \\
\end{array}
\right) 
\end{equation} 
for the longitudinal part of the propagators. The two sectors of the 
propagators can be split since we employ the projectors $P_{\mu\nu}(p)$ and
$L_{\mu\nu}(p)$ which satisfy  
\begin{equation}
\eta_{\mu\nu} ~=~ P_{\mu\nu}(p) ~+~ L_{\mu\nu}(p) ~~~,~~~
P_\mu^{~\nu}(p) L_{\nu\sigma}(p) ~=~ 0 ~.
\end{equation}
We use a similar decomposition for the colour tensors with  
\begin{eqnarray}
{\cal D}^{cdpq}_\xi &=& {\cal D}_\xi \delta^{cp} \delta^{dq} ~+~ {\cal J}_\xi 
f^{cpe} f^{dqe} ~+~ {\cal K}_\xi f^{cde} f^{pqe} ~+~ 
{\cal L}_\xi d_A^{cdpq} ~+~ {\cal M}_\xi \delta^{cd} \delta^{pq} ~+~ 
{\cal N}_\xi \delta^{cq} \delta^{dp} \nonumber \\ 
{\cal D}^{cdpq}_\rho &=& {\cal D}_\rho \delta^{cp} \delta^{dq} ~+~ 
{\cal J}_\rho f^{cpe} f^{dqe} ~+~ {\cal K}_\rho f^{cde} f^{pqe} ~+~ 
{\cal L}_\rho d_A^{cdpq} ~+~ {\cal M}_\rho \delta^{cd} \delta^{pq} ~+~ 
{\cal N}_\rho \delta^{cq} \delta^{dp} \nonumber \\ 
\end{eqnarray} 
and 
\begin{eqnarray}
{\cal D}^{L\, cdpq}_\xi &=& {\cal D}^L_\xi \delta^{cp} \delta^{dq} ~+~ 
{\cal J}^L_\xi f^{cpe} f^{dqe} ~+~ {\cal K}^L_\xi f^{cde} f^{pqe} ~+~ 
{\cal L}^L_\xi d_A^{cdpq} ~+~ {\cal M}^L_\xi \delta^{cd} \delta^{pq} ~+~ 
{\cal N}^L_\xi \delta^{cq} \delta^{dp} \nonumber \\ 
{\cal E}^{L\, cdpq}_\xi &=& {\cal E}^L_\xi \delta^{cp} \delta^{dq} ~+~ 
{\cal F}^L_\xi f^{cpe} f^{dqe} ~+~ {\cal G}^L_\xi f^{cde} f^{pqe} ~+~ 
{\cal H}^L_\xi d_A^{cdpq} ~+~ {\cal Y}^L_\xi \delta^{cd} \delta^{pq} ~+~ 
{\cal Z}^L_\xi \delta^{cq} \delta^{dp} \nonumber \\ 
{\cal E}^{L\, cdpq}_\rho &=& {\cal E}^L_\rho \delta^{cp} \delta^{dq} ~+~ 
{\cal F}^L_\rho f^{cpe} f^{dqe} ~+~ {\cal G}^L_\rho f^{cde} f^{pqe} ~+~ 
{\cal H}^L_\rho d_A^{cdpq} ~+~ {\cal Y}^L_\rho \delta^{cd} \delta^{pq} ~+~ 
{\cal Z}^L_\rho \delta^{cq} \delta^{dp} ~. \nonumber \\ 
\end{eqnarray} 
The inversions in the two sectors proceed via 
\begin{equation}
\Lambda^{\{ab|cd\}} \Pi^{\{cd|pq\}} ~=~ \left(
\begin{array}{ccc}
\delta^{cp} & 0 & 0 \\
0 & \delta^{cp} \delta^{dq} & 0 \\
0 & 0 & \delta^{cp} \delta^{dq} \\
\end{array}
\right) 
\end{equation} 
and
\begin{equation}
\Lambda^{L\,\{ab|cd\}} \Pi^{L\,\{cd|pq\}} ~=~ \left(
\begin{array}{ccc}
\delta^{cp} & 0 & 0 \\
0 & \delta^{cp} \delta^{dq} & 0 \\
0 & 0 & \delta^{cp} \delta^{dq} \\
\end{array}
\right) 
\end{equation} 
where the matrix on the right hand side is the unit matrix in this colour
space basis. In order to handle the products of the colour tensors in this
matrix multiplication we recall, \cite{21}, that 
\begin{eqnarray}
d_A^{abpq} d_A^{cdpq} &=& a_1 \delta^{ab} \delta^{cd} ~+~
a_2 \left( \delta^{ac} \delta^{bd} ~+~ \delta^{ad} \delta^{bc} \right) ~+~
a_3 \left( f^{ace} f^{bde} ~+~ f^{ade} f^{bce} \right) ~+~
a_4 d_A^{abcd} \nonumber \\
f^{ape} f^{bqe} d_A^{cdpq} &=& b_1 \delta^{ab} \delta^{cd} ~+~
b_2 \left( \delta^{ac} \delta^{bd} ~+~ \delta^{ad} \delta^{bc} \right)
\nonumber \\
&& +~ b_3 \left( f^{ace} f^{bde} ~+~ f^{ade} f^{bce} \right) ~+~
b_4 d_A^{abcd} 
\label{decomps}
\end{eqnarray}
where 
\begin{eqnarray}
a_1 &=& -~ \left[ 540 C_A^2 \NA (\NA-3) d_A^{abcd} d_A^{cdpq} d_A^{abpq}
+ 144 (2\NA+19) \left( d_A^{abcd} d_A^{abcd} \right)^2 \right. \nonumber \\
&& \left. ~~~~-~ 150 C_A^4 \NA (3\NA+11) d_A^{abcd} d_A^{abcd}
+ 625 C_A^8 \NA^2 \right] \nonumber \\
&& ~~~ \times ~
\frac{1}{54\NA(\NA-3) [ 12 (\NA + 2) d_A^{efgh} d_A^{efgh} - 25 C_A^4 \NA]}
\nonumber \\
a_2 &=& \left[ 144 (11\NA - 8) \left( d_A^{abcd} d_A^{abcd} \right)^2 \right.
- 1080 C_A^2 \NA (\NA - 3) d_A^{abcd} d_A^{cdpq} d_A^{abpq} \nonumber \\
&& \left. ~+~ 625 C_A^8 \NA^2 - 3000 C_A^4 \NA d_A^{abcd} d_A^{abcd} \right]
\nonumber \\
&& \times ~
\frac{1}{108\NA(\NA-3) [ 12 (\NA + 2) d_A^{efgh} d_A^{efgh} - 25 C_A^4 \NA]}
\nonumber \\
a_3 &=& \frac{[ 12 (\NA + 2) d_A^{abcd} d_A^{abcd} - 25 C_A^4 \NA]}
{54 C_A \NA (\NA-3)} \nonumber \\
a_4 &=& \frac{[ 216 (\NA + 2) d_A^{abcd} d_A^{cdpq} d_A^{abpq} - 125 C_A^6 \NA
- 360 C_A^2 d_A^{abcd} d_A^{abcd}]}{18[ 12 (\NA + 2) d_A^{efgh} d_A^{efgh}
- 25 C_A^4 \NA]}
\end{eqnarray}
and 
\begin{equation}
b_1 ~=~ -~ 2 b_2 ~=~ \frac{[5C_A^4 \NA - 12 d_A^{abcd} d_A^{abcd}]}
{9 C_A \NA (\NA - 3)} ~~,~~
b_3 ~=~ \frac{[6 (\NA-1) d_A^{abcd} d_A^{abcd} - 5 C_A^4 \NA]}
{9 C_A^2 \NA (\NA - 3)} ~~,~~ b_4 ~=~ \frac{C_A}{3} ~. 
\end{equation}

It is clear that the set of linear algebraic equations resulting in multiplying
out the matrices of colour amplitudes will be very involved. For reference in 
the case of the absence of any conventional mass terms these are formally given
in \cite{21,45}. Indeed retaining all possible masses $\mu_i^2$ will be very
complicated and we have provided the propagators for this situation in Appendix
A for the specific case of $SU(3)$. Instead it seems more instructive to 
consider the effect {\em one} particular mass term has on the propagators in 
turn. The motivation for this is to see which masses can produce propagator 
behaviour akin to that observed in lattice simulations. We note that in solving
the set for the longitudinal sector it is important one follows a specific 
algorithm. This is because we are working in the Landau gauge but in order to 
carry out the inversion correctly we must retain a non-zero $\alpha$ at the 
outset. The Landau gauge propagators are deduced at the end by setting 
$\alpha$~$=$~$0$ which will produce a transverse gluon propagator in all cases.
As an aid we note that in the absence of (\ref{brstgen}) and (\ref{opaa}) the 
non-zero propagators are 
\begin{eqnarray}
\langle A^a_\mu(p) A^b_\nu(-p) \rangle &=& -~ 
\frac{\delta^{ab}p^2}{[(p^2)^2+C_A\gamma^4]} P_{\mu\nu}(p) ~-~ 
\frac{\alpha\delta^{ab}p^2}{[(p^2)^2+\alpha C_A\gamma^4]} L_{\mu\nu}(p) 
\nonumber \\
\langle A^a_\mu(p) \xi^{bc}_\nu(-p) \rangle &=& 
\frac{i f^{abc}\gamma^2}{[(p^2)^2+C_A\gamma^4]} P_{\mu\nu}(p) ~+~ 
\frac{i \alpha f^{abc}\gamma^2}{[(p^2)^2+ \alpha C_A\gamma^4]} L_{\mu\nu}(p) 
\nonumber \\
\langle A^a_\mu(p) \rho^{bc}_\nu(-p) \rangle &=& 0 \nonumber \\ 
\langle \xi^{ab}_\mu(p) \xi^{cd}_\nu(-p) \rangle &=& -~ 
\frac{\delta^{ac}\delta^{bd}}{p^2}\eta_{\mu\nu} ~+~
\frac{f^{abe}f^{cde}\gamma^4}{p^2[(p^2)^2+C_A\gamma^4]} P_{\mu\nu}(p) ~+~
\frac{\alpha f^{abe}f^{cde}\gamma^4}{p^2[(p^2)^2+\alpha C_A\gamma^4]} 
L_{\mu\nu}(p) \nonumber \\ 
\langle \xi^{ab}_\mu(p) \rho^{cd}_\nu(-p) \rangle &=& 0 \nonumber \\ 
\langle \rho^{ab}_\mu(p) \rho^{cd}_\nu(-p) \rangle &=& 
\langle \omega^{ab}_\mu(p) \bar{\omega}^{cd}_\nu(-p) \rangle ~=~ -~ 
\frac{\delta^{ac}\delta^{bd}}{p^2} \eta_{\mu\nu} 
\label{propal}
\end{eqnarray} 
which are clearly non-singular in the limit to the Landau gauge.

Given these considerations we record the propagators for each of the masses 
$\mu_i^2$ being non-zero in succession. We append a subscript $i$ to the 
propagators themselves to keep a track of each channel. First, before looking 
at the six localizing ghost possibilities, including the gluon mass we find
\begin{eqnarray}
\langle A^a_\mu(p) A^b_\nu(-p) \rangle_{{\cal X}} &=& -~ 
\frac{\delta^{ab}p^2}{[(p^2)^2+\mu_{{\cal X}}^2p^2+C_A\gamma^4]} P_{\mu\nu}(p) 
\nonumber \\
\langle A^a_\mu(p) \xi^{bc}_\nu(-p) \rangle_{{\cal X}} &=& 
\frac{i f^{abc}\gamma^2}{[(p^2)^2+\mu_{{\cal X}}^2p^2+C_A\gamma^4]} 
P_{\mu\nu}(p) \nonumber \\
\langle A^a_\mu(p) \rho^{bc}_\nu(-p) \rangle_{{\cal X}} &=& 0 \nonumber \\ 
\langle \xi^{ab}_\mu(p) \xi^{cd}_\nu(-p) \rangle_{{\cal X}} &=& -~ 
\frac{\delta^{ac}\delta^{bd}}{p^2}\eta_{\mu\nu} ~+~
\frac{f^{abe}f^{cde}\gamma^4}{p^2[(p^2)^2+\mu_{{\cal X}}^2p^2+C_A\gamma^4]} 
P_{\mu\nu}(p) \nonumber \\ 
\langle \xi^{ab}_\mu(p) \rho^{cd}_\nu(-p) \rangle_{{\cal X}} &=& 0 \nonumber \\ 
\langle \rho^{ab}_\mu(p) \rho^{cd}_\nu(-p) \rangle_{{\cal X}} &=& 
\langle \omega^{ab}_\mu(p) \bar{\omega}^{cd}_\nu(-p) \rangle_{{\cal X}} ~=~ -~ 
\frac{\delta^{ac}\delta^{bd}}{p^2} \eta_{\mu\nu} ~. 
\label{propx}
\end{eqnarray} 
This produces a Stingl propagator which has been observed before, \cite{18}, 
and which will be a common feature in other sets of propagators. However, the 
gluon propagator is suppressed similarly to the original Gribov propagator. For
the Bose ghost there are several massless poles. It is these features of 
suppression and location of massless poles which is the central focus of this 
propagator analysis. If instead we include the mass terms already proposed in 
\cite{36,37,38} we reproduce those results but record them in our current 
notation for completeness. We have 
\begin{eqnarray}
\langle A^a_\mu(p) A^b_\nu(-p) \rangle_{{\cal Q}} &=& -~ 
\frac{\delta^{ab}[p^2+\mu_{{\cal Q}}^2]}
{[(p^2)^2+\mu_{{\cal Q}}^2p^2+C_A\gamma^4]} P_{\mu\nu}(p) \nonumber \\
\langle A^a_\mu(p) \xi^{bc}_\nu(-p) \rangle_{{\cal Q}} &=& 
\frac{i f^{abc}\gamma^2}{[(p^2)^2+\mu_{{\cal Q}}^2p^2+C_A\gamma^4]} 
P_{\mu\nu}(p) \nonumber \\
\langle A^a_\mu(p) \rho^{bc}_\nu(-p) \rangle_{{\cal Q}} &=& 0 \nonumber \\ 
\langle \xi^{ab}_\mu(p) \xi^{cd}_\nu(-p) \rangle_{{\cal Q}} &=& -~ 
\frac{\delta^{ac}\delta^{bd}}{[p^2+\mu_{{\cal Q}}^2]}\eta_{\mu\nu} ~+~
\frac{f^{abe}f^{cde}\gamma^4}
{[p^2+\mu_{{\cal Q}}^2][(p^2)^2+\mu_{{\cal Q}}^2p^2+C_A\gamma^4]} 
P_{\mu\nu}(p) \nonumber \\ 
\langle \xi^{ab}_\mu(p) \rho^{cd}_\nu(-p) \rangle_{{\cal Q}} &=& 0 \nonumber \\ 
\langle \rho^{ab}_\mu(p) \rho^{cd}_\nu(-p) \rangle_{{\cal Q}} &=& 
\langle \omega^{ab}_\mu(p) \bar{\omega}^{cd}_\nu(-p) \rangle_{{\cal Q}} ~=~ -~ 
\frac{\delta^{ac}\delta^{bd}}{[p^2+\mu_{{\cal Q}}^2]} \eta_{\mu\nu} ~. 
\label{propq}
\end{eqnarray} 
Here the gluon propagator does not vanish at zero momentum and there are no
massless poles in any of the colour channels. It was in part this gluon 
propagator freezing which justified examining the condensation of the 
associated BRST invariant operator originally, \cite{36,37,38}. As the 
expressions for both the separate ${\cal W}$ and ${\cal S}$ channel propagators
are cumbersome we have recorded them in Appendix B for the case of $SU(N_c)$. 
However, the case of the ${\cal R}$ channel is similar to that for ${\cal Q}$ 
since 
\begin{eqnarray}
\langle A^a_\mu(p) A^b_\nu(-p) \rangle_{{\cal R}} &=& -~ 
\frac{\delta^{ab}[p^2+\mu_{{\cal R}}^2]}
{[(p^2)^2+\mu_{{\cal R}}^2p^2+C_A\gamma^4]} P_{\mu\nu}(p) \nonumber \\
\langle A^a_\mu(p) \xi^{bc}_\nu(-p) \rangle_{{\cal R}} &=& 
\frac{i f^{abc}\gamma^2}{[(p^2)^2+\mu_{{\cal R}}^2p^2+C_A\gamma^4]} 
P_{\mu\nu}(p) \nonumber \\
\langle A^a_\mu(p) \rho^{bc}_\nu(-p) \rangle_{{\cal R}} &=& 0 \nonumber \\ 
\langle \xi^{ab}_\mu(p) \xi^{cd}_\nu(-p) \rangle_{{\cal R}} &=& -~ 
\frac{\delta^{ac}\delta^{bd}}{p^2}\eta_{\mu\nu} ~+~
\frac{f^{abe}f^{cde}[\mu_{{\cal R}}^2p^2+C_A\gamma^4]}
{C_Ap^2[(p^2)^2+\mu_{{\cal R}}^2p^2+C_A\gamma^4]} P_{\mu\nu}(p) \nonumber \\
&& +~ \frac{f^{abe}f^{cde}\mu_{{\cal R}}^2}{C_Ap^2[p^2+\mu_{{\cal R}}^2]} 
L_{\mu\nu}(p) \nonumber \\ 
\langle \xi^{ab}_\mu(p) \rho^{cd}_\nu(-p) \rangle_{{\cal R}} &=& 0 \nonumber \\ 
\langle \rho^{ab}_\mu(p) \rho^{cd}_\nu(-p) \rangle_{{\cal R}} &=& 
\langle \omega^{ab}_\mu(p) \bar{\omega}^{cd}_\nu(-p) \rangle_{{\cal R}} ~=~ -~ 
\frac{\delta^{ac}\delta^{bd}}{p^2} \eta_{\mu\nu} ~+~ 
\frac{f^{abe}f^{cde}\mu_{{\cal R}}^2}{C_Ap^2[p^2+\mu_{{\cal R}}^2]} 
\eta_{\mu\nu}
\label{propr}
\end{eqnarray} 
and we note that the inclusion of the factor of $C_A$ in (\ref{brstgen}) eases
comparison. Like (\ref{propq}) the gluon propagator freezes. However, the main
difference is that there are massless poles in, for instance, the 
$\xi^{ab}_\mu$ propagator. Whilst it was these massless poles which became
enhanced when the gap equation was satisfied by $\gamma$ in the pure
Gribov-Zwanziger Lagrangian we will show later that there is no similar
enhancement in this case. The situation for the ${\cal P}$ channel has 
parallels with the previous set since
\begin{eqnarray}
\langle A^a_\mu(p) A^b_\nu(-p) \rangle_{{\cal P}} &=& -~ 
\frac{\delta^{ab}p^2}{[(p^2)^2+C_A\gamma^4]} P_{\mu\nu}(p) \nonumber \\
\langle A^a_\mu(p) \xi^{bc}_\nu(-p) \rangle_{{\cal P}} &=& 
\frac{i f^{abc}\gamma^2}{[(p^2)^2+C_A\gamma^4]} P_{\mu\nu}(p) 
\nonumber \\
\langle A^a_\mu(p) \rho^{bc}_\nu(-p) \rangle_{{\cal P}} &=& 0 \nonumber \\ 
\langle \xi^{ab}_\mu(p) \xi^{cd}_\nu(-p) \rangle_{{\cal P}} &=& -~ 
\frac{\delta^{ac}\delta^{bd}}{p^2}\eta_{\mu\nu} ~+~
\frac{f^{abe}f^{cde}\gamma^4}{p^2[(p^2)^2+C_A\gamma^4]} P_{\mu\nu}(p) ~+~
\frac{\delta^{ab}\delta^{cd}\mu_{{\cal P}}^2}{\NA p^2[p^2+\mu_{{\cal P}}^2]} 
\eta_{\mu\nu} \nonumber \\ 
\langle \xi^{ab}_\mu(p) \rho^{cd}_\nu(-p) \rangle_{{\cal P}} &=& 0 \nonumber \\ 
\langle \rho^{ab}_\mu(p) \rho^{cd}_\nu(-p) \rangle_{{\cal P}} &=& 
\langle \omega^{ab}_\mu(p) \bar{\omega}^{cd}_\nu(-p) \rangle_{{\cal P}} ~=~ -~ 
\frac{\delta^{ac}\delta^{bd}}{p^2} \eta_{\mu\nu} ~+~  
\frac{\delta^{ab}\delta^{cd}\mu_{{\cal P}}^2}{\NA p^2[p^2+\mu_{{\cal P}}^2]} 
\eta_{\mu\nu} 
\label{propp}
\end{eqnarray} 
producing massless poles but gluon suppression instead of freezing. By contrast
the ${\cal T}$ channel has gluon freezing but no massless poles because 
\begin{eqnarray}
\langle A^a_\mu(p) A^b_\nu(-p) \rangle_{{\cal T}} &=& -~ 
\frac{\delta^{ab}[p^2-\mu_{{\cal T}}^2]}
{[(p^2)^2-\mu_{{\cal T}}^2p^2+C_A\gamma^4]} P_{\mu\nu}(p) \nonumber \\
\langle A^a_\mu(p) \xi^{bc}_\nu(-p) \rangle_{{\cal T}} &=& 
\frac{i f^{abc}\gamma^2}{[(p^2)^2-\mu_{{\cal T}}^2p^2+C_A\gamma^4]} 
P_{\mu\nu}(p) \nonumber \\
\langle A^a_\mu(p) \rho^{bc}_\nu(-p) \rangle_{{\cal T}} &=& 0 \nonumber \\ 
\langle \xi^{ab}_\mu(p) \xi^{cd}_\nu(-p) \rangle_{{\cal T}} &=& -~ 
\frac{\delta^{ac}\delta^{bd}p^2}{[(p^2)^2-\mu_{{\cal T}}^4]}\eta_{\mu\nu} ~+~
\frac{f^{abe}f^{cde}\gamma^4}
{[p^2-\mu_{{\cal T}}^2][(p^2)^2-\mu_{{\cal T}}^2p^2+C_A\gamma^4]} 
P_{\mu\nu}(p) \nonumber \\
&& +~ \frac{\delta^{ad}\delta^{bc}\mu_{{\cal T}}^2}{[(p^2)^2-\mu_{{\cal T}}^4]} 
\eta_{\mu\nu} \nonumber \\ 
\langle \xi^{ab}_\mu(p) \rho^{cd}_\nu(-p) \rangle_{{\cal T}} &=& 0 \nonumber \\ 
\langle \rho^{ab}_\mu(p) \rho^{cd}_\nu(-p) \rangle_{{\cal T}} &=& 
\langle \omega^{ab}_\mu(p) \bar{\omega}^{cd}_\nu(-p) \rangle_{{\cal T}} ~=~ -~ 
\frac{\delta^{ac}\delta^{bd}}{[p^2-\mu_{{\cal T}}^2]} \eta_{\mu\nu} ~+~ 
\frac{\delta^{ad}\delta^{bc}\mu_{{\cal T}}^2}{[(p^2)^2-\mu_{{\cal T}}^4]} 
\eta_{\mu\nu} ~. 
\label{propt}
\end{eqnarray} 
Though, of the sets we have recorded this appears to be the one which is least
likely to be realistic since a type of tachyonic pole appears which would
violate causality. 

As there is interest in effective gluon masses we record the propagators for
two more general situations. Those with a non-zero $\mu_{{\cal X}}^2$ and 
either a non-zero $\mu_{{\cal Q}}^2$ or a non-zero $\mu_{{\cal R}}^2$. The 
former case has been discussed in \cite{36} but as we will concentrate on the 
${\cal R}$ channel we will examine the consequences of a gluon mass for this 
case briefly as well. The motivation for a non-zero $\mu_{{\cal X}}^2$ 
originates from the same argument as that leading to (\ref{condop}). If we 
integrate the gluon propagator then one obtains a non-zero vacuum expectation 
value for $\half A^a_\mu A^{a\,\mu}$. For completeness and to allow us to 
contrast with other situations we note that the ${\cal XQ}$ case is 
\begin{eqnarray}
\langle A^a_\mu(p) A^b_\nu(-p) \rangle_{{\cal XQ}} &=& -~ 
\frac{\delta^{ab}[p^2+\mu_{{\cal Q}}^2]}
{[(p^2)^2+(\mu_{{\cal X}}^2+\mu_{{\cal Q}}^2)p^2
+\mu_{{\cal X}}^2\mu_{{\cal Q}}^2+C_A\gamma^4]} P_{\mu\nu}(p) \nonumber \\
\langle A^a_\mu(p) \xi^{bc}_\nu(-p) \rangle_{{\cal XQ}} &=& 
\frac{i f^{abc}\gamma^2}
{[(p^2)^2+(\mu_{{\cal X}}^2+\mu_{{\cal Q}}^2)p^2
+\mu_{{\cal X}}^2\mu_{{\cal Q}}^2+C_A\gamma^4]} P_{\mu\nu}(p) 
\nonumber \\
\langle A^a_\mu(p) \rho^{bc}_\nu(-p) \rangle_{{\cal XQ}} &=& 0 \nonumber \\ 
\langle \xi^{ab}_\mu(p) \xi^{cd}_\nu(-p) \rangle_{{\cal XQ}} &=& -~ 
\frac{\delta^{ac}\delta^{bd}}{[p^2+\mu_{{\cal Q}}^2]}\eta_{\mu\nu} \nonumber \\
&& +~ \frac{f^{abe}f^{cde} \gamma^4}
{[p^2+\mu_{{\cal Q}}^2][(p^2)^2+(\mu_{{\cal X}}^2+\mu_{{\cal Q}}^2)p^2
+\mu_{{\cal X}}^2\mu_{{\cal Q}}^2+C_A\gamma^4]} P_{\mu\nu}(p) \nonumber \\ 
\langle \xi^{ab}_\mu(p) \rho^{cd}_\nu(-p) \rangle_{{\cal XQ}} &=& 0
\nonumber \\ 
\langle \rho^{ab}_\mu(p) \rho^{cd}_\nu(-p) \rangle_{{\cal XQ}} &=& 
\langle \omega^{ab}_\mu(p) \bar{\omega}^{cd}_\nu(-p) \rangle_{{\cal XQ}} ~=~ -~ 
\frac{\delta^{ac}\delta^{bd}}{[p^2+\mu_{{\cal Q}}^2]} \eta_{\mu\nu} ~. 
\label{propxq}
\end{eqnarray} 
The presence of the non-zero $\mu_{{\cal X}}^2$ does not alter the properties
significantly from the pure ${\cal Q}$ channel case. There is still gluon
freezing and no massless poles in the full set. The structure of the Stingl
propagator is unsurprisingly affected. Equally when $\mu_{{\cal X}}^2$ and 
$\mu_{{\cal R}}^2$ are both non-zero the propagators behave in essence in the 
same way as the ${\cal R}$ channel ones since 
\begin{eqnarray}
\langle A^a_\mu(p) A^b_\nu(-p) \rangle_{{\cal XR}} &=& -~ 
\frac{\delta^{ab}[p^2+\mu_{{\cal R}}^2]}
{[(p^2)^2+(\mu_{{\cal X}}^2+\mu_{{\cal R}}^2)p^2
+\mu_{{\cal X}}^2\mu_{{\cal R}}^2+C_A\gamma^4]} P_{\mu\nu}(p) \nonumber \\
\langle A^a_\mu(p) \xi^{bc}_\nu(-p) \rangle_{{\cal XR}} &=& 
\frac{i f^{abc}\gamma^2}
{[(p^2)^2+(\mu_{{\cal X}}^2+\mu_{{\cal R}}^2)p^2
+\mu_{{\cal X}}^2\mu_{{\cal R}}^2+C_A\gamma^4]} P_{\mu\nu}(p) 
\nonumber \\
\langle A^a_\mu(p) \rho^{bc}_\nu(-p) \rangle_{{\cal XR}} &=& 0 \nonumber \\ 
\langle \xi^{ab}_\mu(p) \xi^{cd}_\nu(-p) \rangle_{{\cal XR}} &=& -~ 
\frac{\delta^{ac}\delta^{bd}}{p^2}\eta_{\mu\nu} ~+~
\frac{f^{abe}f^{cde}[\mu_{{\cal R}}^2p^2+\mu_{{\cal X}}^2\mu_{{\cal R}}^2
+C_A\gamma^4]}
{C_Ap^2[(p^2)^2+(\mu_{{\cal X}}^2+\mu_{{\cal R}}^2)p^2
+\mu_{{\cal X}}^2\mu_{{\cal R}}^2+C_A\gamma^4]} P_{\mu\nu}(p) \nonumber \\ 
&& +~ \frac{f^{abe}f^{cde}\mu_{{\cal R}}^2}{C_Ap^2[p^2+\mu_{{\cal R}}^2]} 
L_{\mu\nu}(p) \nonumber \\ 
\langle \xi^{ab}_\mu(p) \rho^{cd}_\nu(-p) \rangle_{{\cal XR}} &=& 0
\nonumber \\ 
\langle \rho^{ab}_\mu(p) \rho^{cd}_\nu(-p) \rangle_{{\cal XR}} &=& 
\langle \omega^{ab}_\mu(p) \bar{\omega}^{cd}_\nu(-p) \rangle_{{\cal XR}} ~=~ -~ 
\frac{\delta^{ac}\delta^{bd}}{p^2} \eta_{\mu\nu} ~+~ 
\frac{f^{abe}f^{cde}\mu_{{\cal R}}^2}{C_Ap^2[p^2+\mu_{{\cal R}}^2]} 
L_{\mu\nu}(p) ~. \nonumber \\ 
\label{propxr}
\end{eqnarray} 
Clearly there is gluon suppression together with massless poles. For both
(\ref{propxq}) and (\ref{propxr}) one could have the situation where if
$\mu_{{\cal X}}^2$~$=$~$-$~$\mu_{{\cal Q}}^2$ or
$\mu_{{\cal X}}^2$~$=$~$-$~$\mu_{{\cal R}}^2$ then one would return to a Gribov
type denominator in the gluon propagator. However, the main point to appreciate
from this analysis is that there is another possibility of obtaining a frozen 
gluon via the condensation of a BRST invariant dimension two operator which has
a different colour structure to that considered in \cite{36,37,38}. This 
derives from (\ref{propr}) and we will refer to it as the ${\cal R}$ channel 
mechanism for the moment and study it in more detail in the next section.

Prior to that we need to briefly discuss the renormalization of the general 
operator ${\cal O}^{abcd}$ since we will be performing loop computations. It 
has already been shown that the contraction with $\delta^{ac} \delta^{bd}$ 
produces a renormalizable operator, \cite{36,37,38}. However, the more general 
operator is clearly renormalizable multiplicatively in the Landau gauge by the 
same reasoning. To partly verify this and check our internal conventions we 
have renormalized ${\cal O}^{abcd}$ to two loops by inserting the operator in 
either a $\phi^{ab}_\mu$ or a $\omega^{ab}_\mu$ $2$-point function where the 
momentum flows in through one of the legs and out through the operator itself. 
Our momentum configuration allows one to apply the {\sc Mincer} algorithm of 
\cite{48} but recoded, \cite{49}, in the symbolic manipulation language 
{\sc Form}, \cite{50}. As the two external fields in the Green's function each 
carry one Lorentz index then we have to contract all the Feynman integrals with
the tensor $\eta_{\mu\nu}/d$ so that {\sc Mincer} can be applied to a Lorentz 
scalar. The factor of $d$ in the denominator arises from the normalization. 
Concerning the colour indices we do no projection on the four free indices of 
the operators which allows us to check that the divergent part of the Green's 
function is only associated with the colour structure of the original Feynman 
rule for the operator. In essence this is how we check that the renormalization
of the general operator is multiplicative. For the Green's function with 
$\phi^{ab}_\mu$ legs there are $4$ one loop and $160$ two loop Feynman 
diagrams. The respective figures for the $\omega^{ab}_\mu$ Green's function are
$1$ and $25$. The diagrams are generated by the {\sc Qgraf} package, \cite{51},
before being converted into {\sc Form} input notation. Then the {\sc Form} 
version of the {\sc Mincer} algorithm, \cite{49}, is applied. As we are using 
an automatic Feynman diagram computation procedure we use the method of 
rescaling of bare quantities such as the coupling constant in order to deduce 
the overall final renormalization constant for the Green's functions we 
renormalize, \cite{52}. For both Green's functions we find that the operator is
multiplicatively renormalized and with the {\em same} renormalization constant 
$Z_{\cal O}$. Not unexpectedly like the earlier colour contraction the more 
general operator ${\cal O}^{abcd}$ satisfies the Slavnov-Taylor identity
\begin{equation} 
Z_\phi Z_{\cal O} ~=~ 1
\end{equation}
to two loops in $\MSbar$ where $Z_\phi$ is the renormalization constant of the 
original localizing ghost field $\phi^{ab}_\mu$. The renormalization constants 
for the real and imaginary parts of this field $\phi^{ab}_\mu$ are equivalent. 

\sect{${\cal R}$ channel mass.}

Having considered the different forms the propagators of the spin-$1$ sector
can take when there are a variety of single mass terms originating from the 
BRST invariant operator ${\cal O}^{abcd}$, we now focus on the ${\cal R}$
sector mass in detail. Like the ${\cal Q}$ case it has a frozen gluon
propagator. The aim is to give evidence that there is a dynamical origin for 
such a mass as the previous analysis merely assumed the existence of such
additional mass terms. The approach we take is the same as that of \cite{37}
for the ${\cal Q}$ channel. Indeed given the similarities between the 
${\cal Q}$ and ${\cal R}$ colour contractions of ${\cal O}^{abcd}$ a large 
amount of the results of \cite{37} can immediately be transferred to the
present case without detailed re-analysis. For instance, the inclusion of the
general operator (\ref{brstgen}) does not destroy the renormalizability of the
pure Gribov-Zwanziger Lagrangian, (\ref{laggz}). Moreover, we use the same
procedure to examine the dynamical origin of an ${\cal R}$ channel mass which
is the construction of an effective potential for the operator. From this it
will transpire that there is a non-zero vacuum expectation value for the 
operator which therefore condenses to produce the mass term we analysed
previously. However, it turns out that from the way we have set up the
${\cal R}$ term of (\ref{opaa}) that the effective potential is formally the 
{\em same} as that for the ${\cal Q}$ case whence we can merely translate all
those results to the present case. For instance, the starting point of the
application of the local composite operator formalism, \cite{39,40,41,42}, to 
the operator is the one loop energy functional $W[J]$ where $J$ is the source 
coupling to the ${\cal R}$ channel operator. Summing up the conventional set of
contributing one loop diagrams produces
\begin{equation}
W[J] ~=~ -~ \frac{d\NA\gamma^4}{2g^2} ~+~ 
\frac{d\NA\zeta J\gamma^2}{g^2} ~+~
\frac{(d-1)\NA}{2} \int_k \ln \left[ k^2 \left[ k^2 
+ \frac{C_A\gamma^4}{[k^2+J]} \right] \right] ~+~ O(g^2)
\end{equation}
where $\zeta$ is the local composite operator parameter, \cite{39,40,41,42}.
Clearly this is formally similar to that for the ${\cal Q}$ channel and hence 
we merely recall the subsequent properties. Defining  
\begin{equation}
\sigma(x) ~=~ \frac{\delta W[J]}{\delta J(x)}
\end{equation}
then the effective action, $\Gamma[\sigma]$, is constructed from the Legendre 
transformation
\begin{equation}
\Gamma[\sigma] ~=~ W[J] ~-~ \int d^4x \, J(x) \sigma(x) ~.
\end{equation}
Setting
\begin{equation}
\sigma(x) ~=~ \sigma_0 ~+~ \hat{\sigma}(x)
\end{equation}
where
\begin{equation}
\sigma_0 ~=~ \frac{d\NA \zeta \gamma^2}{g^2}
\end{equation} 
then we find the same non-zero value for the condensate as before, \cite{37},
\begin{equation}
\left. \frac{}{} \hat{\sigma} \right|_{J=0} ~=~ -~ 
\frac{3\NA\sqrt{C_A}\gamma^2}{64\pi} ~.
\end{equation}
With this observation then the corresponding subsequent analysis of \cite{37}
in respect of the gluon propagator will hold for the ${\cal R}$ channel mass. 
This is due to the fact that the gluon propagators of (\ref{propq}) and 
(\ref{propr}) are formally the same. Therefore, the estimate for the value 
where the gluon propagator freezes, which is in qualitative agreement with the 
lattice, applies in this case too.

Given this complete parallel between the two cases it might appear that there
is no justification in posing an alternative way of having a frozen gluon in a 
refinement of the Gribov-Zwanziger formalism. However, there is a key 
difference and it resides in the Bose ghost sector. From (\ref{propq}) and
(\ref{propr}) there are massless poles in the $\xi^{ab}_\mu$ propagator for the
latter case but not for the former. Recently, Zwanziger has argued, \cite{43},
that in the pure case there is an enhancement of the Bose ghost in the infrared
which is a non-perturbative property of the theory. The argument is based on 
the spontaneous breaking of the BRST symmetry in a theory where fields are 
constrained by the horizon condition. The structure of the propagator has been 
examined at one loop in the $\MSbar$ scheme in the zero momentum limit and an 
enhanced $\xi^{ab}_\mu$ emerges, \cite{21}. There is also enhancement in the 
$\rho^{ab}_\mu$ case too. In essence the starting point for this is the 
massless poles of the original propagators, (\ref{prop}). However, one needs to
compute the one loop corrections to all the $2$-point functions of the fields
in the zero momentum limit and apply the gap equation satisfied by $\gamma$.
Then the leading momentum term vanishes to ensure that the resultant 
propagators have a dipole behaviour in the infrared as opposed to the canonical
behaviour of a massless field. The analysis for $\xi^{ab}_\mu$ is hampered by
the colour tensor structure of the $2$-point functions and hence only certain
colour channels of its propagator enhance. 

Therefore, given that the structure of the $\xi^{ab}_\mu$ propagators in
(\ref{propq}) and (\ref{propr}) are different it is worth examining the 
structure of the ${\cal R}$ propagators in the infrared limit. However, as the
gap equation is important for this we need to compute the one loop expression
for $\gamma$. This is achieved by evaluating (\ref{hordefloc}) using the mixed
propagator of (\ref{propr}). As the denominator factors are of a Stingl form
in order to evaluate the basic Feynman integrals we need to first rewrite 
the factor formally in a more conventional fashion. Therefore, we define two
additional mass parameters, $\mu_\pm^2$, given by 
\begin{equation}
\mu_{{\cal R}}^2 ~=~ \mu_+^2 ~+~ \mu_-^2 ~~~,~~~ 
C_A \gamma^4 ~=~ \mu_+^2 \mu_-^2 ~.
\end{equation}
This choice is motivated by the relation
\begin{equation}
(p^2)^2 + \mu_{{\cal R}}^2 p^2 + C_A\gamma^4 ~=~ [ p^2 + \mu_+^2 ] 
[ p^2 + \mu_-^2 ] ~.
\end{equation}
Solving the relations produces the mapping 
\begin{eqnarray}
\mu_+^2 &=& \frac{1}{2} \left[ \mu_{{\cal R}}^2 ~+~ 
\sqrt{ \mu_{{\cal R}}^4 - 4 C_A \gamma^4 } \right] \nonumber \\ 
\mu_-^2 &=& \frac{1}{2} \left[ \mu_{{\cal R}}^2 ~-~ 
\sqrt{ \mu_{{\cal R}}^4 - 4 C_A \gamma^4 } \right] ~.
\end{eqnarray}
With this factorization it is a straightforward matter to determine the
$\MSbar$ gap equation and we find 
\begin{equation}
1 ~=~ C_A \left[ \frac{5}{8} ~-~ \frac{3}{8} \ln \left[
\frac{C_A\gamma^4}{\mu^4} \right] ~-~
\frac{3\mu_{{\cal R}}^2}{8\sqrt{\mu_{{\cal R}}^4 - 4 C_A \gamma^4}} \ln \left[
\frac{\mu_+^2}{\mu_-^2} \right] \right] a ~+~ O(a^2) ~.
\label{gapr}
\end{equation}
In the limit of $\mu_{{\cal R}}^2$~$\rightarrow$~$0$ one recovers the gap
equation for $\gamma$ of the pure Gribov-Zwanziger case. This expression is
formally the same as that for the ${\cal Q}$ channel, \cite{36,37}, if one 
simply replaces $\mu_{{\cal R}}^2$ by $\mu_{{\cal Q}}^2$. This is not 
unexpected since the mixed propagators are formally the same at this order. At 
two loops one would expect the gap equations to be different merely because at 
that order the other spin-$1$ propagators will appear in the Feynman diagrams.  

Before considering the infrared structure of the $\xi^{ab}_\mu$ propagator we
first examine the Faddeev-Popov ghost propagator. This is partly to highlight
the technique one follows but in a case which is not complicated by the colour
tensor structure as well as to verify the loss of ghost enhancement which has 
to be a feature of the ${\cal R}$ channel in order to be consistent with the
evidence from the lattice. However, the situation for the behaviour of the 
Faddeev-Popov ghost propagator in the infrared is the same as that for 
${\cal Q}$ since at one loop only one diagram contributes to the $2$-point 
function. Defining the ghost propagator in terms of its form factor, 
$D_c(p^2)$, by
\begin{equation}
\langle c^a(p) \bar{c}^b(-p) \rangle ~=~ \frac{D_c(p^2)}{p^2} \delta^{ab} 
\end{equation}
we have 
\begin{eqnarray}
D_c(p^2) &=& -~ \left[ 1 ~-~ C_A \left[ \frac{5}{8} ~-~ \frac{3}{8} \ln \left[
\frac{C_A\gamma^4}{\mu^4} \right] ~+~
\frac{3\mu_{{\cal R}}^2}{8\sqrt{\mu_{{\cal R}}^4 - 4 C_A \gamma^4}} \ln \left[
\frac{\mu_+^2}{\mu_-^2} \right] \right. \right. \nonumber \\
&& \left. \left. ~~~~~~~~~~~~~~~~~~+~ \left[ \left[ \frac{1}{8} \ln \left[ 
\frac{(p^2)^2}{C_A\gamma^4} \right] - \frac{11}{24}
+ \frac{1}{8} \sqrt{\mu_{{\cal R}}^4 - 4 C_A \gamma^4} \ln \left[
\frac{\mu_+^2}{\mu_-^2} \right] 
\right] \frac{\mu_{{\cal R}}^2}{C_A\gamma^4} \right. \right. \right. 
\nonumber \\
&& \left. \left. \left. ~~~~~~~~~~~~~~~~~~~~~~~~~+~ 
\frac{1}{4\sqrt{\mu_{{\cal R}}^4 - 4 C_A \gamma^4}} \ln \left[
\frac{\mu_+^2}{\mu_-^2} \right] \right] p^2 ~+~ O\left( (p^2)^2 \right) 
\right] a \right. \nonumber \\
&& \left. ~~~~~+~ O(a^2) \right]^{-1} 
\end{eqnarray}
in the limit as $p^2$~$\rightarrow$~$0$ where we have included the $O(p^2)$
contribution. Ordinarily this term would govern the infrared behaviour of the
propagator in the infrared as the leading order term in momentum would vanish
when $\gamma$ satisfies the gap equation. That does not happen due to the
$O(\mu_{{\cal R}}^2)$ term which has the opposite sign to the analogous term
which appears in the gap equation, (\ref{gapr}). Therefore, as with the
${\cal Q}$ channel there is no Faddeev-Popov ghost enhancement for ${\cal R}$
either. Again this is in keeping with the lattice data which sees a minor
variation from $1/p^2$ behaviour of the ghost propagator in the infrared,
\cite{23,24,25,26,27,28,29,30,31,32,34}. As the colour structure of the 
$\rho^{ab}_\mu$ and $\omega^{ab}_\mu$ propagators is equally as trivial as that
of the Faddeev-Popov ghost then the propagators of these fields are not 
enhanced either. 

It is interesting to trace the origin of the sign discrepancy that prevents the
enhancement happening at one loop. For instance, the ghost $2$-point function 
correction is purely from one Feynman diagram and that includes a gluon 
propagator. By contrast evaluating the horizon vacuum expectation value 
involves the mixed propagator at one loop. Therefore, from examining these 
propagators it transpires that it is the part of the gluon propagator leading 
to its freezing which is responsible for the difference in the signs of the 
respective terms in the gap equation and the ghost $2$-point function. In other
words if the $\mu_{{\cal R}}^2$ term in the gluon propagator numerator was 
absent then there would be one loop ghost enhancement even with the Stingl type
propagator whose denominator has to be the {\em same} in both the gluon and 
mixed propagator. With this observation it is simple to determine which of the 
various channels we have introduced in an earlier section will contain enhanced
Faddeev-Popov propagators. This will happen for ${\cal X}$, ${\cal P}$ and
${\cal S}$. So, interestingly, a massive gluon but with a Gribov width would be 
infrared suppressed whilst satisfying the Kugo-Ojima confinement criterion at 
one loop in this context. Whether there is enhancement in various colour 
channels for the $\xi^{ab}_\mu$ fields and if so which ones, is more 
complicated to determine since it would require the structure of the 
$\xi^{ab}_\mu$ $2$-point function as well as all the other $2$-point functions.

To illustrate the complexity of such an exercise even in a simple case we 
return to the ${\cal R}$ channel and record the explicit one loop form factors 
for all the spin-$1$ field $2$-point functions. We have
\begin{eqnarray}
{\cal X} &=& -~ \left[ p^2 ~-~ \left[ C_A \left[ 
\frac{27\mu_{{\cal R}}^2 C_A\gamma^4}{32[\mu_{{\cal R}}^4-4C_A\gamma^4]} ~-~ 
\frac{27C_A^2\gamma^8\sqrt{\mu_{{\cal R}}^4-4C_A\gamma^4}}
{16[\mu_{{\cal R}}^4-4C_A\gamma^4]^2}
\ln \left[ \frac{\mu_+^2}{\mu_-^2} \right] \right. \right. \right. \nonumber \\
&& \left. \left. \left. ~~~~~~~~~~~~~~~~~~~~~+~ \frac{3}{8} 
\sqrt{\mu_{{\cal R}}^4-4C_A\gamma^4} \ln \left[ \frac{\mu_+^2}{\mu_-^2} 
\right] ~+~ \frac{3\mu_{{\cal R}}^2}{8} \ln \left[ 
\frac{C_A\gamma^4}{\mu_{{\cal R}}^4} \right] 
\right. \right. \right. \nonumber \\ 
&& \left. \left. \left. ~~~~~~~~~~~~~~~~~~~~~+~ 
\frac{5\mu_{{\cal R}}^2p^2}{192C_A\gamma^4} 
\left[ \sqrt{\mu_{{\cal R}}^4-4C_A\gamma^4} \ln \left[ \frac{\mu_+^2}{\mu_-^2}
\right] ~+~ \mu_{{\cal R}}^2 \ln \left[ \frac{C_A\gamma^4}{\mu_{{\cal R}}^4} 
\right] \right] \right. \right. \right. \nonumber \\ 
&& \left. \left. \left. ~~~~~~~~~~~~~~~~~~~~~+~ C_A \gamma^4 p^2 \left[  
\frac{143\mu_{{\cal R}}^2\sqrt{\mu_{{\cal R}}^4-4C_A\gamma^4}}
{64[\mu_{{\cal R}}^4-4C_A\gamma^4]^2} \ln \left[ \frac{\mu_+^2}{\mu_-^2} 
\right] ~-~ \frac{73}{16[\mu_{{\cal R}}^4-4C_A\gamma^4]} \right] 
\right. \right. \right. \nonumber \\ 
&& \left. \left. \left. ~~~~~~~~~~~~~~~~~~~~~+~ C_A^2 \gamma^8 p^2 \left[  
\frac{9\mu_{{\cal R}}^2\sqrt{\mu_{{\cal R}}^4-4C_A\gamma^4}}
{16[\mu_{{\cal R}}^4-4C_A\gamma^4]^3} \ln \left[ \frac{\mu_+^2}{\mu_-^2} 
\right] ~-~ \frac{9}{8[\mu_{{\cal R}}^4-4C_A\gamma^4]^2} \right] 
\right. \right. \right. \nonumber \\ 
&& \left. \left. \left. ~~~~~~~~~~~~~~~~~~~~~+~ p^2 \left[  
\frac{457\mu_{{\cal R}}^2\sqrt{\mu_{{\cal R}}^4-4C_A\gamma^4}}
{384[\mu_{{\cal R}}^4-4C_A\gamma^4]} \ln \left[ \frac{\mu_+^2}{\mu_-^2} 
\right] ~-~ \frac{7}{96} ~-~ \frac{121}{128} \ln \left[ 
\frac{C_A\gamma^4}{\mu_{{\cal R}}^4} \right] 
\right. \right. \right. \right. \nonumber \\ 
&& \left. \left. \left. \left. ~~~~~~~~~~~~~~~~~~~~~~~~~~~~~-~ 
\frac{25}{12} \ln \left[ \frac{\mu_{{\cal R}}^2}{\tilde{\mu}^2} \right] ~-~ 
\frac{1}{12} \ln \left[ \frac{p^2}{\tilde{\mu}^2} \right] \right] \right] 
\right. \right. \nonumber \\ 
&& \left. \left. ~~~~~~~~~~~~~~~+~ p^2 T_F \Nf \left[ \frac{4}{3} \ln \left[ 
\frac{p^2}{\tilde{\mu}^2} \right] ~-~ \frac{20}{9} \right] \right] a
\right] ~+~ O \left( (p^2)^2 \right) \nonumber \\
{\cal U} &=& i \gamma^2 \left[ 1 ~+~ \left[ C_A \left[ 
\frac{5\mu_{{\cal R}}^2p^2}{192C_A\gamma^4} 
\left[ \sqrt{\mu_{{\cal R}}^4-4C_A\gamma^4} \ln \left[ \frac{\mu_+^2}{\mu_-^2}
\right] ~+~ \mu_{{\cal R}}^2 \ln \left[ \frac{C_A\gamma^4}{\mu_{{\cal R}}^4} 
\right] \right] \right. \right. \right. \nonumber \\ 
&& \left. \left. \left. ~~~~~~~~~~~~~~~~~~~~+~  
\frac{5C_A \gamma^4 p^2 \sqrt{\mu_{{\cal R}}^4-4C_A\gamma^4}}
{4[\mu_{{\cal R}}^4-4C_A\gamma^4]^2} \ln \left[ \frac{\mu_+^2}{\mu_-^2} \right] 
\right. \right. \right. \nonumber \\ 
&& \left. \left. \left. ~~~~~~~~~~~~~~~~~~~~+~ p^2 \left[  
\frac{71\sqrt{\mu_{{\cal R}}^4-4C_A\gamma^4}}
{96[\mu_{{\cal R}}^4-4C_A\gamma^4]} \ln \left[ \frac{\mu_+^2}{\mu_-^2} 
\right] ~-~ \frac{5\mu_{{\cal R}}^2}{8[\mu_{{\cal R}}^4-4C_A\gamma^4]} \right] 
\right] \right] a \right] \nonumber \\
&& +~ O \left( (p^2)^2 \right) \nonumber \\ 
{\cal V} &=& O(a^2) \nonumber \\ 
{\cal Q}_\xi &=& -~ \left[ 1 ~-~ C_A \left[ \frac{5}{8} ~-~ \frac{3}{8} 
\ln \left[ \frac{C_A\gamma^4}{\mu^4} \right] ~+~
\frac{3\mu_{{\cal R}}^2}{8\sqrt{\mu_{{\cal R}}^4 - 4 C_A \gamma^4}} \ln \left[
\frac{\mu_+^2}{\mu_-^2} \right] \right] a \right] p^2 ~+~ 
O\left( (p^2)^2 \right) \nonumber \\
{\cal W}_\xi &=& \left[ 
\frac{7C_A\gamma^4}{36[\mu_{{\cal R}}^4-4C_A\gamma^4]} ~-~ 
\frac{7\mu_{{\cal R}}^2C_A\gamma^4\sqrt{\mu_{{\cal R}}^4-4C_A\gamma^4}}
{72[\mu_{{\cal R}}^4-4C_A\gamma^4]^2} \ln \left[ \frac{\mu_+^2}{\mu_-^2} 
\right] \right. \nonumber \\
&& \left. ~-~ 
\frac{\mu_{{\cal R}}^2 \sqrt{\mu_{{\cal R}}^4-4C_A\gamma^4}}
{8[\mu_{{\cal R}}^4-4C_A\gamma^4]} \ln \left[ \frac{\mu_+^2}{\mu_-^2} \right] 
\right] a p^2 ~+~ O \left( (p^2)^2 \right) \nonumber \\ 
{\cal R}_\xi &=& \mu_{{\cal R}}^2 ~+~ \left[ 
\frac{11 C_A\gamma^4}{72[\mu_{{\cal R}}^4-4C_A\gamma^4]} ~-~ 
\frac{11\mu_{{\cal R}}^2C_A\gamma^4\sqrt{\mu_{{\cal R}}^4-4C_A\gamma^4}}
{144[\mu_{{\cal R}}^4-4C_A\gamma^4]^2} \ln \left[ \frac{\mu_+^2}{\mu_-^2} 
\right] \right. \nonumber \\
&& \left. ~~~~~~~~~~~-~ 
\frac{\mu_{{\cal R}}^2 \sqrt{\mu_{{\cal R}}^4-4C_A\gamma^4}}
{8[\mu_{{\cal R}}^4-4C_A\gamma^4]}
\ln \left[ \frac{\mu_+^2}{\mu_-^2} \right] \right] a p^2 ~+~ 
O \left( (p^2)^2 \right) \nonumber \\ 
{\cal S}_\xi &=& \left[ \frac{7\gamma^4}{6[\mu_{{\cal R}}^4-4C_A\gamma^4]} ~-~ 
\frac{7\mu_{{\cal R}}^2\gamma^4\sqrt{\mu_{{\cal R}}^4-4C_A\gamma^4}}
{12[\mu_{{\cal R}}^4-4C_A\gamma^4]^2} \ln \left[ \frac{\mu_+^2}{\mu_-^2} 
\right] \right. \nonumber \\
&& \left. ~-~ \frac{3\mu_{{\cal R}}^2 \sqrt{\mu_{{\cal R}}^4-4C_A\gamma^4}}
{4C_A[\mu_{{\cal R}}^4-4C_A\gamma^4]} \ln \left[ \frac{\mu_+^2}{\mu_-^2} 
\right] \right] a p^2 ~+~ O \left( (p^2)^2 \right) \nonumber \\ 
{\cal P}_\xi &=& {\cal T}_\xi ~=~ {\cal P}^L_\xi ~=~ {\cal T}^L_\xi ~=~ O(a^2)
\nonumber \\
{\cal X}^L &=& -~ \left[ \frac{p^2}{\alpha} ~-~ \left[ C_A \left[ 
\frac{27\mu_{{\cal R}}^2 C_A\gamma^4}{32[\mu_{{\cal R}}^4-4C_A\gamma^4]} ~-~ 
\frac{27C_A^2\gamma^8\sqrt{\mu_{{\cal R}}^4-4C_A\gamma^4}}
{16[\mu_{{\cal R}}^4-4C_A\gamma^4]^2} \ln \left[ \frac{\mu_+^2}{\mu_-^2} 
\right] \right. \right. \right. \nonumber \\
&& \left. \left. \left. ~~~~~~~~~~~~~~~~~~~~~~+~ \frac{3}{8} 
\sqrt{\mu_{{\cal R}}^4-4C_A\gamma^4} \ln \left[ \frac{\mu_+^2}{\mu_-^2} 
\right] ~+~ \frac{3\mu_{{\cal R}}^2}{8} \ln \left[ 
\frac{C_A\gamma^4}{\mu_{{\cal R}}^4} \right] 
\right. \right. \right. \nonumber \\ 
&& \left. \left. \left. ~~~~~~~~~~~~~~~~~~~~~~-~ 
\frac{5\mu_{{\cal R}}^2p^2}{64C_A\gamma^4} 
\left[ \sqrt{\mu_{{\cal R}}^4-4C_A\gamma^4} \ln \left[ \frac{\mu_+^2}{\mu_-^2}
\right] ~+~ \mu_{{\cal R}}^2 \ln \left[ \frac{C_A\gamma^4}{\mu_{{\cal R}}^4} 
\right] \right] \right. \right. \right. \nonumber \\ 
&& \left. \left. \left. ~~~~~~~~~~~~~~~~~~~~~~-~ 
\frac{9\mu_{{\cal R}}^2C_A\gamma^4 p^2\sqrt{\mu_{{\cal R}}^4-4C_A\gamma^4}}
{64[\mu_{{\cal R}}^4-4C_A\gamma^4]^2} \ln \left[ \frac{\mu_+^2}{\mu_-^2} 
\right] 
\right. \right. \right. \nonumber \\ 
&& \left. \left. \left. ~~~~~~~~~~~~~~~~~~~~~~+~ C_A^2 \gamma^8 p^2 \left[  
\frac{27\mu_{{\cal R}}^2\sqrt{\mu_{{\cal R}}^4-4C_A\gamma^4}}
{16[\mu_{{\cal R}}^4-4C_A\gamma^4]^3} \ln \left[ \frac{\mu_+^2}{\mu_-^2} 
\right] ~-~ 
\frac{27}{8[\mu_{{\cal R}}^4-4C_A\gamma^4]^2} \right] 
\right. \right. \right. \nonumber \\ 
&& \left. \left. \left. ~~~~~~~~~~~~~~~~~~~~~~+~ p^2 \left[  
\frac{11\mu_{{\cal R}}^2\sqrt{\mu_{{\cal R}}^4-4C_A\gamma^4}}
{128[\mu_{{\cal R}}^4-4C_A\gamma^4]} \ln \left[ \frac{\mu_+^2}{\mu_-^2} 
\right] ~+~ \frac{29}{96} ~+~ 
\frac{63}{128} \ln \left[ \frac{C_A\gamma^4}{\mu_{{\cal R}}^4} \right] 
\right. \right. \right. \right. \nonumber \\ 
&& \left. \left. \left. \left. ~~~~~~~~~~~~~~~~~~~~~~~~~~~~~~-~ 
\frac{1}{4} \ln \left[ \frac{p^2}{\mu_{{\cal R}}^2} \right] \right] 
\right] \right] a \right] ~+~ O \left( (p^2)^2 \right) \nonumber \\
{\cal U}^L &=& i \gamma^2 \left[ 1 ~+~ \left[ C_A \left[ 
\frac{p^2}{64C_A\gamma^4} 
\left[ \sqrt{\mu_{{\cal R}}^4-4C_A\gamma^4} \ln \left[ \frac{\mu_+^2}{\mu_-^2}
\right] ~+~ \mu_{{\cal R}}^2 \ln \left[ \frac{C_A\gamma^4}{\mu_{{\cal R}}^4} 
\right] \right] \right. \right. \right. \nonumber \\ 
&& \left. \left. \left. ~~~~~~~~~~~~~~~~~~~~-~ 
\frac{3C_A \gamma^4 p^2 \sqrt{\mu_{{\cal R}}^4-4C_A\gamma^4}}
{16[\mu_{{\cal R}}^4-4C_A\gamma^4]^2} \ln \left[ \frac{\mu_+^2}{\mu_-^2} 
\right] \right. \right. \right. \nonumber \\ 
&& \left. \left. \left. ~~~~~~~~~~~~~~~~~~~~+~ p^2 \left[  
\frac{\sqrt{\mu_{{\cal R}}^4-4C_A\gamma^4}}{8[\mu_{{\cal R}}^4-4C_A\gamma^4]}
\ln \left[ \frac{\mu_+^2}{\mu_-^2} \right] ~+~ 
\frac{3\mu_{{\cal R}}^2}{32[\mu_{{\cal R}}^4-4C_A\gamma^4]} \right] 
\right] \right] a \right] \nonumber \\
&& +~ O \left( (p^2)^2 \right) \nonumber \\ 
{\cal V}^L &=& -~ i \gamma^2 \left[ C_A \left[ 
\frac{3p^2}{64C_A\gamma^4} 
\left[ \sqrt{\mu_{{\cal R}}^4-4C_A\gamma^4} \ln \left[ \frac{\mu_+^2}{\mu_-^2}
\right] ~+~ \mu_{{\cal R}}^2 \ln \left[ \frac{C_A\gamma^4}{\mu_{{\cal R}}^4} 
\right] \right] \right. \right. \nonumber \\ 
&& \left. \left. ~~~~~~~~~~~~~~~+~ 
\frac{3p^2\sqrt{\mu_{{\cal R}}^4-4C_A\gamma^4}}
{8[\mu_{{\cal R}}^4-4C_A\gamma^4]} \ln \left[ \frac{\mu_+^2}{\mu_-^2} \right] 
\right] \right] a ~+~ O \left( (p^2)^2 \right) \nonumber \\ 
{\cal Q}_\xi &=& {\cal Q}^L_\xi ~+~ O(a^2) ~=~ {\cal Q}_\rho ~+~ O(a^2) ~=~ 
{\cal Q}^L_\rho ~+~ O(a^2) \nonumber \\ 
{\cal W}^L_\xi &=& \left[ 
\frac{5 C_A\gamma^4}{12[\mu_{{\cal R}}^4-4C_A\gamma^4]} ~-~ 
\frac{5\mu_{{\cal R}}^2C_A\gamma^4\sqrt{\mu_{{\cal R}}^4-4C_A\gamma^4}}
{24[\mu_{{\cal R}}^4-4C_A\gamma^4]^2} \ln \left[ \frac{\mu_+^2}{\mu_-^2} 
\right] \right. \nonumber \\
&& \left. ~~-~ 
\frac{\mu_{{\cal R}}^2 \sqrt{\mu_{{\cal R}}^4-4C_A\gamma^4}}
{8[\mu_{{\cal R}}^4-4C_A\gamma^4]} \ln \left[ \frac{\mu_+^2}{\mu_-^2} \right] 
\right] a p^2 ~+~ O \left( (p^2)^2 \right) \nonumber \\ 
{\cal R}^L_\xi &=& \mu_{{\cal R}}^2 ~-~ \left[ 
\frac{5 C_A\gamma^4}{24[\mu_{{\cal R}}^4-4C_A\gamma^4]} ~-~ 
\frac{5\mu_{{\cal R}}^2C_A\gamma^4\sqrt{\mu_{{\cal R}}^4-4C_A\gamma^4}}
{48[\mu_{{\cal R}}^4-4C_A\gamma^4]^2} \ln \left[ \frac{\mu_+^2}{\mu_-^2} 
\right] \right. \nonumber \\
&& \left. ~~~~~~~~~~~+~ 
\frac{\mu_{{\cal R}}^2 \sqrt{\mu_{{\cal R}}^4-4C_A\gamma^4}}
{8[\mu_{{\cal R}}^4-4C_A\gamma^4]} \ln \left[ \frac{\mu_+^2}{\mu_-^2} \right] 
\right] a p^2 ~+~ O \left( (p^2)^2 \right) \nonumber \\ 
{\cal S}^L_\xi &=& \left[ 
\frac{5\gamma^4}{2[\mu_{{\cal R}}^4-4C_A\gamma^4]} ~-~ 
\frac{5\mu_{{\cal R}}^2\gamma^4\sqrt{\mu_{{\cal R}}^4-4C_A\gamma^4}}
{4[\mu_{{\cal R}}^4-4C_A\gamma^4]^2} \ln \left[ \frac{\mu_+^2}{\mu_-^2} \right]
\right. \nonumber \\
&& \left. ~-~ \frac{3\mu_{{\cal R}}^2 \sqrt{\mu_{{\cal R}}^4-4C_A\gamma^4}}
{4C_A[\mu_{{\cal R}}^4-4C_A\gamma^4]} \ln \left[ \frac{\mu_+^2}{\mu_-^2} 
\right] \right] a p^2 ~+~ O \left( (p^2)^2 \right) \nonumber \\
{\cal W}_\rho &=& {\cal W}^L_\rho ~=~ -~ 
\frac{\mu_{{\cal R}}^2 \sqrt{\mu_{{\cal R}}^4-4C_A\gamma^4}}
{8[\mu_{{\cal R}}^4-4C_A\gamma^4]} \ln \left[ \frac{\mu_+^2}{\mu_-^2} 
\right] a p^2 ~+~ O \left( (p^2)^2 \right) \nonumber \\ 
{\cal R}_\rho &=& {\cal R}^L_\rho ~=~ \mu_{{\cal R}}^2 ~-~ 
\frac{\mu_{{\cal R}}^2 \sqrt{\mu_{{\cal R}}^4-4C_A\gamma^4}}
{8[\mu_{{\cal R}}^4-4C_A\gamma^4]} \ln \left[ \frac{\mu_+^2}{\mu_-^2} 
\right] a p^2 ~+~ O \left( (p^2)^2 \right) \nonumber \\ 
{\cal S}_\rho &=& {\cal S}^L_\rho ~=~ -~ 
\frac{3\mu_{{\cal R}}^2 \sqrt{\mu_{{\cal R}}^4-4C_A\gamma^4}}
{4C_A[\mu_{{\cal R}}^4-4C_A\gamma^4]} \ln \left[ \frac{\mu_+^2}{\mu_-^2} 
\right] a p^2 ~+~ O \left( (p^2)^2 \right) 
\end{eqnarray}
where $\tilde{\mu}$ is the mass scale introduced to ensure that the coupling
constant remains dimensionless in $d$-dimensions as we are using dimensional
regularization. The divergences have been removed by the $\MSbar$ scheme
prescription. As expected ${\cal Q}_\xi$ is similar to the Faddeev-Popov ghost 
$2$-point function. In the pure case this function was solely responsible for 
producing the overall enhancement of the $\xi^{ab}_\mu$ propagator observed in 
\cite{21} which was consistent with the general analysis of \cite{43}. However,
since the leading term of the momentum expansion does not vanish when the gap 
equation is satisfied this leads to the absence of enhancement for 
$\xi^{ab}_\mu$ in the ${\cal R}$ channel. To be more specific we find the 
leading behaviour in the $p^2$~$\rightarrow$~$0$ limit is 
\begin{eqnarray}
\langle \xi^{ab}_\mu(p) \xi^{cd}_\nu(-p) \rangle_{\cal R} & \sim & 
\frac{1}{2{\cal Q}_0 p^2 a} \left[ \delta^{ac} \delta^{bd} ~-~
\delta^{ad} \delta^{bc} ~-~ \frac{2}{C_A} f^{abe} f^{cde} \right] \eta_{\mu\nu}
\nonumber \\ 
\langle \rho^{ab}_\mu(p) \rho^{cd}_\nu(-p) \rangle_{\cal R} & \sim & 
\frac{1}{2{\cal Q}_0 p^2 a} \left[ \delta^{ac} \delta^{bd} ~-~ 
\delta^{ad} \delta^{bc} ~-~ \frac{2}{C_A} f^{abe} f^{cde} \right] \eta_{\mu\nu}
\label{proprir}
\end{eqnarray} 
where
\begin{equation}
{\cal Q}_0 ~=~ \frac{1}{4\sqrt{\mu_{{\cal R}}^4 - 4 C_A \gamma^4}} \ln \left[
\frac{\mu_+^2}{\mu_-^2} \right] ~+~ \left[ \frac{1}{8} 
\sqrt{\mu_{{\cal R}}^4 - 4 C_A \gamma^4} \ln \left[ \frac{\mu_+^2}{\mu_-^2} 
\right] ~-~ \frac{1}{8} \ln \left[ \frac{C_A\gamma^4}{(p^2)^2} \right] ~-~ 
\frac{11}{24} \right]  \frac{\mu_{{\cal R}}^2}{C_A\gamma^4} 
\end{equation}
which is derived using the standard procedure from the coefficient of the 
leading term in the zero momentum limit {\em after} the gap equation has been 
set. It is the same as that for the Faddeev-Popov ghost. As expected there is 
no enhancement and unlike the pure Gribov-Zwanziger case the leading momentum 
behaviour involves a logarithm of the momentum. However, the Bose ghost 
propagators diverge in the same way as the Faddeev-Popov ghost. Interestingly 
the colour tensor structure is the same as that for the enhanced $\xi^{ab}_\mu$
propagator of the pure Gribov-Zwanziger case discussed in \cite{43,21} and we 
recall the leading behaviour of both Bose ghosts in that case is 
\begin{eqnarray}
\langle \xi^{ab}_\mu(p) \xi^{cd}_\nu(-p) \rangle & \sim & 
\frac{4 \gamma^2}{\pi \sqrt{C_A} (p^2)^2 a} \left[ \delta^{ad} \delta^{bc}
- \delta^{ac} \delta^{bd} \right] \eta_{\mu\nu} ~+~ 
\frac{8 \gamma^2}{\pi C_A^{3/2} (p^2)^2 a} f^{abe} f^{cde} 
P_{\mu\nu}(p) \nonumber \\ 
\langle \rho^{ab}_\mu(p) \rho^{cd}_\nu(-p) \rangle & \sim & -~ 
\frac{8 \gamma^2}{\pi \sqrt{C_A} (p^2)^2 a} \delta^{ac} \delta^{bd} 
\eta_{\mu\nu} ~. 
\label{propir}
\end{eqnarray} 
So far in our discussions we have noted that the properties of the ${\cal Q}$
and ${\cal R}$ channel masses are the same. However, (\ref{proprir}) clearly
represents the first departure from that similarity. For the ${\cal Q}$ case 
all colour channels of the $\xi^{ab}_\mu$ propagator freeze to finite values in
the zero momentum limit. By contrast the ${\cal R}$ channel $\xi^{ab}_\mu$ 
propagator neither freezes nor enhances. Instead its behaviour is in essence 
the same as that of the Faddeev-Popov ghost since the coefficient of the 
leading term is the same but differs in the colour tensor structure. That the 
same tensor structure as (\ref{propir}) emerges at leading order for both cases
is a consequence of the peculiarities of the inversion of the matrix of colour 
structures. However, the adjoint colour projection that Zwanziger focused on in
\cite{43} will not enhance nor diverge but freeze to an non-zero value.

There are several main consequences of our ${\cal R}$ channel analysis. First,
it appears that there are now two possibilities of modelling lattice data by
refining the Gribov-Zwanziger Lagrangian with {\em one} additional mass 
operator. Both the ${\cal Q}$ and ${\cal R}$ channels reproduce gluon freezing 
to a non-zero value and ghost non-enhancement. However, to determine which one 
is correct would require a lattice computation of the $\xi^{ab}_\mu$ propagator
in the zero momentum limit. It would seem to us that this would be a 
non-trivial exercise since the lattice gauge fixing procedure used pays no 
attention to a field, $\xi^{ab}_\mu$, which only appears in the 
Gribov-Zwanziger Lagrangian and is necessary to localize the non-local horizon 
operator. However, one could consider instead the correlation of a non-local 
projection of the gluon field itself since the equation of motion from 
(\ref{laggz})
\begin{equation}
\left( \partial^\nu D_\nu \xi_\mu \right)^{ab} ~=~ i \gamma^2 f^{abc} A^c_\mu
\label{xieom}
\end{equation}
produces the relation
\begin{equation}
\xi^{ab}_\mu ~=~ i \gamma^2 \left( \frac{1}{\partial^\nu D_\nu} \right)^{ad} 
f^{dbc} A^c_\mu 
\label{xieominv}
\end{equation}
where we have been careful in including the colour indices of the inverse
Faddeev-Popov operator to correct an error in earlier work, \cite{20,21}. Again
this would appear to open up other difficulties since the presence of a 
non-locality via the Faddeev-Popov operator may be hard to define in a discrete
spacetime in an unambiguous way. Aside from that it may even be restrictive 
both financially and computationally to produce accurate enough data in order 
to determine the behaviour at zero momentum definitively. Currently, the state 
of the art to obtain the gluon propagator behaviour at zero momentum involves a 
new formulation of the lattice definition of the linear covariant gauge, 
\cite{53}. Another problem relates to whether (\ref{xieominv}) is the proper 
definition of the field whose correlator is the relevant object to examine. For
instance, (\ref{xieominv}) assumes there is no mass term initially for the 
$\xi^{ab}_\mu$ field and that the mass term has a dynamical origin. However, if
one has the mass term present then it could be argued that (\ref{xieom}) should
be replaced by something such as  
\begin{equation}
\left( \partial^\nu D_\nu \xi_\mu \right)^{ab} ~+~ 
\frac{\mu_{{\cal R}}^2}{C_A} f^{abe} f^{cde} \xi^{cd}_\mu ~=~ i \gamma^2 
f^{abc} A^c_\mu ~. 
\end{equation}
Though in the infrared limit this would effectively be the same as an adjoint 
projection of the gluon. So it is not clear in this case whether this would 
allow one to make a clear statement on the behaviour of the propagators we have
considered, $\langle \xi^{ab}_\mu(p) \xi^{cd}_\nu(-p) \rangle_i$ and $\langle 
\rho^{ab}_\mu(p) \rho^{cd}_\nu(-p) \rangle_i$. 

\sect{Discussion.}

There are several main features arising out of our analysis which was to
attempt to model lattice data with a more {\em natural} perturbation of the 
original Gribov-Zwanziger Lagrangian with a different BRST invariant dimension 
two operator compared to \cite{36,37,38}. First, if one is to reconcile the 
lattice and Schwinger-Dyson observations that the Landau gauge gluon propagator
freezes to a non-zero value and the Faddeev-Popov ghost does not enhance, by 
modelling with a refined Gribov-Zwanziger Lagrangian, then it turns out that
there is more than one way to do this. The original analysis of \cite{36,37,38}
used only one colour projection but it has been shown here that the alternative
${\cal R}$ channel projection can equally accommodate a non-zero frozen gluon 
propagator and unenhanced ghost propagator. If one wished to determine which of
these single operator extensions was consistent with the lattice results then 
the test resides in the infrared behaviour of the propagators of the Bose 
localizing ghost. In the earlier ${\cal Q}$ channel solution the $\xi^{ab}_\mu$
propagator freezes to a non-zero value. By contrast in the ${\cal R}$ channel 
case the propagator of this field has the {\em same} momentum behaviour in the
infrared as the Faddeev-Popov ghost propagator. Given this it might be better 
in future to refer instead to the ${\cal Q}$ channel as the gluon-like massive
or decoupling scenario and that of ${\cal R}$ as the Faddeev-Popov ghost-like 
massive or decoupling solution. In some sense the resolution by numerical work 
may not in fact be possible given the amount of computing resources which would
be required to determine the correlation of a non-local projection of the gluon
field. Moreover, both sets of localizing ghosts are an inherent feature of 
accommodating the original non-local Gribov horizon operator and as such would 
have no parallel or direct concept in a lattice construction. 

We have also noted that the inclusion of the dimension two gluon mass operator,
which is BRST invariant as well, into the propagators for each of the 
${\cal Q}$ and ${\cal R}$ channels does not alter their behaviour from the 
situation when the gluon operator was absent. This in fact opened a wider 
question as to whether all the possible colour projections should not be 
considered simultaneously. As the gluon and Faddeev-Popov ghost propagators are
the only two which have been analysed numerically the fact that the gluon 
freezes to a non-zero value can also be accommodated by contributions from four
different colour channels as can be seen in Appendix A. From a numerical point 
of view the freezing to a non-zero value for the gluon propagator can never 
resolve each of the four different mass scales which arise there. Again only 
data on the remaining spin-$1$ propagators could ever possibly determine this.
It would require, for instance, a substantial amount of numerical fitting of 
all the mass parameters. Though as there are more than seven different 
propagator form factors then this over-redundancy would provide independent
consistency checks on the seven mass parameters where we include the gluon
mass parameter in the counting here. However, we should recall one of our
underlying assumptions in this context. In (\ref{brstgen}) we have taken the
masses for $\rho^{ab}_\mu$, $\xi^{ab}_\mu$ and $\omega^{ab}_\mu$ to be the 
same in order to ensure our additional operator is BRST invariant. If the BRST
symmetry is broken then there is no reason why, for example, any of the masses 
of the localizing fields should be equal. If this were the case then the 
propagators we have discussed throughout would be much more complicated. 
Moreover, this would become apparent in all the different propagator form 
factors analogous to those of Appendix A. Again to test this scenario would 
appear to be computationally impractical for reasons we have already mentioned.
However, one alternative theoretical way to shed some light on the interplay of
the different colour projections would be to extend the effective potential 
analysis of the single projection case for the ${\cal Q}$ and ${\cal R}$ 
channels to include all seven cases simultaneously. Whilst it is a non-trivial 
task to apply the local composite operator method for this, finding a stable 
absolute minimum of the effective potential would indicate which is the most 
energetically favoured solution or solutions. This is currently in progress. 
However, it seems from earlier experience, with the construction of the 
enhanced Bose ghost propagator in the pure Gribov-Zwanziger Lagrangian, that 
the subtleties of the intricate nature of the general colour tensors of the 
localizing ghosts have a significant effect on the infrared structure of this 
Lagrangian. One only has to recall the decompositions (\ref{decomps}) to see 
how the effective potential construction will mix up colour tensors of the 
operator ${\cal O}^{abcd}$.

Finally, in providing a complete analysis for each of the seven individual
colour projections separately we have noticed several features in the interplay
of a frozen gluon propagator and a non-enhanced Faddeev-Popov ghost propagator 
with the structure of the original propagators when the appropriate gap 
equation is implemented at one loop. Specifically if the propagators of the 
spin-$1$ fields have a massless pole in a colour or Lorentz channel then 
whether it becomes enhanced in the infrared depends on whether the gluon 
propagator freezes to a non-zero value or not. In the case of the original 
Gribov-Zwanziger Lagrangian this was the case as is now evident in the recent 
work of \cite{43,21} for the localizing Bose ghost. Here this particular 
feature emerges in several of the cases. For instance, with a gluon mass 
operator only or in the ${\cal S}$ channel only then when the appropriate 
horizon condition for $\gamma$ is satisfied there appears to be ghost 
enhancement. Whilst there is currently no lattice evidence for this scaling 
scenario since the massive or decoupling solution appears to be favoured by 
many different simulations, \cite{23,24,25,26,27,28,29,30,31,32,34,53}, it is 
perhaps worth noting that if instead a scaling solution had been found then for
this situation the {\em pure} Gribov-Zwanziger structure may not actually have 
been the {\em unique} explanation.

\appendix

\sect{Full propagators for $SU(3)$.}

In this appendix we record the explicit expressions for the propagators for the 
specific colour group $SU(3)$ for all channels together. First, the transverse 
sector is 
\begin{eqnarray}
{\cal A} &=& -~ \left[ 2 \mu_{{\cal Q}}^2 + 2 \mu_{{\cal R}}^2 
- 2 \mu_{{\cal T}}^2 + 3 \mu_{{\cal W}}^2 + 2 p^2 \right] \nonumber \\
&& ~~ \times
\left[ [ 3 \mu_{{\cal W}}^2 + 2 p^2 - 2 \mu_{{\cal T}}^2 + 2 \mu_{{\cal R}}^2
+ 2 \mu_{{\cal Q}}^2 ] [ \mu_{{\cal X}}^2 + p^2 ] + 6 \gamma^4 \right]^{-1}
\nonumber \\
{\cal B} &=& 2 i \gamma^2 
\left[ [ 3 \mu_{{\cal W}}^2 + 2 p^2 + 2 \mu_{{\cal T}}^2 + 2 \mu_{{\cal R}}^2
+ 2 \mu_{{\cal Q}}^2 ] [ \mu_{{\cal X}}^2 + p^2 ] + 6 \gamma^4 \right]^{-1}
\nonumber \\
{\cal D}_\xi &=& {\cal D}_\xi^L ~=~ {\cal D}_\rho ~=~ {\cal D}_\rho^L
\nonumber \\
&=& \frac{1}{2} \left[ -~ 16 (\mu_{{\cal Q}}^2)^3 
- 56 (\mu_{{\cal Q}}^2)^2 \mu_{{\cal S}}^2 
- 32 (\mu_{{\cal Q}}^2)^2 \mu_{{\cal T}}^2 
- 24 (\mu_{{\cal Q}}^2)^2 \mu_{{\cal W}}^2 
- 48 (\mu_{{\cal Q}}^2)^2 p^2 
\right. \nonumber \\
&& \left. ~~~
-~ 54 \mu_{{\cal Q}}^2 (\mu_{{\cal S}}^2)^2 
- 56 \mu_{{\cal Q}}^2 \mu_{{\cal S}}^2 \mu_{{\cal T}}^2 
- 40 \mu_{{\cal Q}}^2 \mu_{{\cal S}}^2 \mu_{{\cal W}}^2 
- 112 \mu_{{\cal Q}}^2 \mu_{{\cal S}}^2 p^2
- 16 \mu_{{\cal Q}}^2 (\mu_{{\cal T}}^2)^2 
\right. \nonumber \\
&& \left. ~~~
-~ 24 \mu_{{\cal Q}}^2 \mu_{{\cal T}}^2 \mu_{{\cal W}}^2 
- 64 \mu_{{\cal Q}}^2 \mu_{{\cal T}}^2 p^2 
+ 16 \mu_{{\cal Q}}^2 (\mu_{{\cal W}}^2)^2 
- 48 \mu_{{\cal Q}}^2 \mu_{{\cal W}}^2 p^2 
- 48 \mu_{{\cal Q}}^2 (p^2)^2 
\right. \nonumber \\
&& \left. ~~~
-~ 9 (\mu_{{\cal S}}^2)^3 
- 6 (\mu_{{\cal S}}^2)^2 \mu_{{\cal T}}^2 
- 21 (\mu_{{\cal S}}^2)^2 \mu_{{\cal W}}^2 
- 54 (\mu_{{\cal S}}^2)^2 p^2 
- 24 \mu_{{\cal S}}^2 \mu_{{\cal T}}^2 \mu_{{\cal W}}^2 
\right. \nonumber \\
&& \left. ~~~
-~ 56 \mu_{{\cal S}}^2 \mu_{{\cal T}}^2 p^2 
- 40 \mu_{{\cal S}}^2 \mu_{{\cal W}}^2 p^2 
- 56 \mu_{{\cal S}}^2 (p^2)^2 
- 16 (\mu_{{\cal T}}^2)^2 p^2 - 24 \mu_{{\cal T}}^2 \mu_{{\cal W}}^2 p^2 
\right. \nonumber \\
&& \left. ~~~
-~ 32 \mu_{{\cal T}}^2 (p^2)^2 + 12 (\mu_{{\cal W}}^2)^3 
+ 16 (\mu_{{\cal W}}^2)^2 p^2 - 24 \mu_{{\cal W}}^2 (p^2)^2 
- 16 (p^2)^3 \right] \nonumber \\
&& \times
\left[ 2 \mu_{{\cal Q}}^2 + 3 \mu_{{\cal S}}^2 + 2 \mu_{{\cal T}}^2 
+ 3 \mu_{{\cal W}}^2 + 2 p^2 \right]^{-1}
\left[ 2 \mu_{{\cal Q}}^2 + 3 \mu_{{\cal S}}^2 + 2 \mu_{{\cal T}}^2 
- 2 \mu_{{\cal W}}^2 + 2 p^2 \right]^{-1} \nonumber \\
&& \times 
\left[ 2 \mu_{{\cal Q}}^2 + \mu_{{\cal S}}^2 + 2 \mu_{{\cal T}}^2 
+ 2 \mu_{{\cal W}}^2 + 2 p^2 \right]^{-1}
\left[ \mu_{{\cal Q}}^2 - \mu_{{\cal T}}^2 + p^2 \right]^{-1} \nonumber \\
{\cal J}_\xi &=& {\cal J}_\xi^L ~=~ {\cal J}_\rho ~=~ {\cal J}_\rho^L 
\nonumber \\
&=& 4 \mu_{{\cal W}}^2
\left[ 2 \mu_{{\cal Q}}^2 + 3 \mu_{{\cal S}}^2 + 2 \mu_{{\cal T}}^2 
+ 3 \mu_{{\cal W}}^2 + 2 p^2 \right]^{-1} 
\left[2 \mu_{{\cal Q}}^2 + 3 \mu_{{\cal S}}^2 + 2 \mu_{{\cal T}}^2 
- 2 \mu_{{\cal W}}^2 + 2 p^2 \right]^{-1} \nonumber \\
{\cal K}_\xi &=& \frac{1}{3} \left[ 24 \gamma^4 (\mu_{{\cal Q}}^2)^2 
+ 72 \gamma^4 \mu_{{\cal Q}}^2 \mu_{{\cal S}}^2 
+ 48 \gamma^4 \mu_{{\cal Q}}^2 \mu_{{\cal T}}^2 
- 24 \gamma^4 \mu_{{\cal Q}}^2 \mu_{{\cal W}}^2 
+ 48 \gamma^4 \mu_{{\cal Q}}^2 p^2 + 54 \gamma^4 (\mu_{{\cal S}}^2)^2 
\right. \nonumber \\
&& \left. ~~~
+~ 72 \gamma^4 \mu_{{\cal S}}^2 \mu_{{\cal T}}^2 
+ 18 \gamma^4 \mu_{{\cal S}}^2 \mu_{{\cal W}}^2 
+ 72 \gamma^4 \mu_{{\cal S}}^2 p^2 + 24 \gamma^4 (\mu_{{\cal T}}^2)^2 
+ 48 \gamma^4 \mu_{{\cal T}}^2 \mu_{{\cal W}}^2 
+ 48 \gamma^4 \mu_{{\cal T}}^2 p^2 
\right. \nonumber \\
&& \left. ~~~
-~ 36 \gamma^4 (\mu_{{\cal W}}^2)^2 
- 24 \gamma^4 \mu_{{\cal W}}^2 p^2 + 24 \gamma^4 (p^2)^2 
+ 8 (\mu_{{\cal Q}}^2)^2 \mu_{{\cal R}}^2 \mu_{{\cal X}}^2 
+ 8 (\mu_{{\cal Q}}^2)^2 \mu_{{\cal R}}^2 p^2 
\right. \nonumber \\
&& \left. ~~~
+~ 24 \mu_{{\cal Q}}^2 \mu_{{\cal R}}^2 \mu_{{\cal S}}^2 \mu_{{\cal X}}^2 
+ 24 \mu_{{\cal Q}}^2 \mu_{{\cal R}}^2 \mu_{{\cal S}}^2 p^2 
+ 16 \mu_{{\cal Q}}^2 \mu_{{\cal R}}^2 \mu_{{\cal T}}^2 \mu_{{\cal X}}^2 
+ 16 \mu_{{\cal Q}}^2 \mu_{{\cal R}}^2 \mu_{{\cal T}}^2 p^2 
\right. \nonumber \\
&& \left. ~~~
-~ 8 \mu_{{\cal Q}}^2 \mu_{{\cal R}}^2 \mu_{{\cal W}}^2 \mu_{{\cal X}}^2 
- 8 \mu_{{\cal Q}}^2 \mu_{{\cal R}}^2 \mu_{{\cal W}}^2 p^2 
+ 16 \mu_{{\cal Q}}^2 \mu_{{\cal R}}^2 \mu_{{\cal X}}^2 p^2 
+ 16 \mu_{{\cal Q}}^2 \mu_{{\cal R}}^2 (p^2)^2 
+ 36 \mu_{{\cal Q}}^2 \mu_{{\cal S}}^2 \mu_{{\cal W}}^2 \mu_{{\cal X}}^2 
\right. \nonumber \\
&& \left. ~~~
+~ 36 \mu_{{\cal Q}}^2 \mu_{{\cal S}}^2 \mu_{{\cal W}}^2 p^2 
+ 48 \mu_{{\cal Q}}^2 \mu_{{\cal T}}^2 \mu_{{\cal W}}^2 \mu_{{\cal X}}^2 
+ 48 \mu_{{\cal Q}}^2 \mu_{{\cal T}}^2 \mu_{{\cal W}}^2 p^2
- 12 \mu_{{\cal Q}}^2 (\mu_{{\cal W}}^2)^2 \mu_{{\cal X}}^2 
\right. \nonumber \\
&& \left. ~~~
-~ 12 \mu_{{\cal Q}}^2 (\mu_{{\cal W}}^2)^2 p^2 
+ 18 \mu_{{\cal R}}^2 (\mu_{{\cal S}}^2)^2 \mu_{{\cal X}}^2 
+ 18 \mu_{{\cal R}}^2 (\mu_{{\cal S}}^2)^2 p^2 
+ 24 \mu_{{\cal R}}^2 \mu_{{\cal S}}^2 \mu_{{\cal T}}^2 \mu_{{\cal X}}^2 
+ 24 \mu_{{\cal R}}^2 \mu_{{\cal S}}^2 \mu_{{\cal T}}^2 p^2 
\right. \nonumber \\
&& \left. ~~~
+~ 6 \mu_{{\cal R}}^2 \mu_{{\cal S}}^2 \mu_{{\cal W}}^2 \mu_{{\cal X}}^2 
+ 6 \mu_{{\cal R}}^2 \mu_{{\cal S}}^2 \mu_{{\cal W}}^2 p^2 
+ 24 \mu_{{\cal R}}^2 \mu_{{\cal S}}^2 \mu_{{\cal X}}^2 p^2 
+ 24 \mu_{{\cal R}}^2 \mu_{{\cal S}}^2 (p^2)^2 
+ 8 \mu_{{\cal R}}^2 (\mu_{{\cal T}}^2)^2 \mu_{{\cal X}}^2 
\right. \nonumber \\
&& \left. ~~~
+~ 8 \mu_{{\cal R}}^2 (\mu_{{\cal T}}^2)^2 p^2 
+ 16 \mu_{{\cal R}}^2 \mu_{{\cal T}}^2 \mu_{{\cal W}}^2 \mu_{{\cal X}}^2 
+ 16 \mu_{{\cal R}}^2 \mu_{{\cal T}}^2 \mu_{{\cal W}}^2 p^2 
+ 16 \mu_{{\cal R}}^2 \mu_{{\cal T}}^2 \mu_{{\cal X}}^2 p^2 
+ 16 \mu_{{\cal R}}^2 \mu_{{\cal T}}^2 (p^2)^2 
\right. \nonumber \\
&& \left. ~~~
-~ 12 \mu_{{\cal R}}^2 (\mu_{{\cal W}}^2)^2 \mu_{{\cal X}}^2 
- 12 \mu_{{\cal R}}^2 (\mu_{{\cal W}}^2)^2 p^2 
- 8 \mu_{{\cal R}}^2 \mu_{{\cal W}}^2 \mu_{{\cal X}}^2 p^2 
- 8 \mu_{{\cal R}}^2 \mu_{{\cal W}}^2 (p^2)^2 
+ 8 \mu_{{\cal R}}^2 \mu_{{\cal X}}^2 (p^2)^2 
\right. \nonumber \\
&& \left. ~~~
+~ 8 \mu_{{\cal R}}^2 (p^2)^3
+ 27 (\mu_{{\cal S}}^2)^2 \mu_{{\cal W}}^2 \mu_{{\cal X}}^2 
+ 27 (\mu_{{\cal S}}^2)^2 \mu_{{\cal W}}^2 p^2 
+ 36 \mu_{{\cal S}}^2 \mu_{{\cal T}}^2 \mu_{{\cal W}}^2 \mu_{{\cal X}}^2 
+ 36 \mu_{{\cal S}}^2 \mu_{{\cal T}}^2 \mu_{{\cal W}}^2 p^2 
\right. \nonumber \\
&& \left. ~~~
+~ 9 \mu_{{\cal S}}^2 (\mu_{{\cal W}}^2)^2 \mu_{{\cal X}}^2 
+ 9 \mu_{{\cal S}}^2 (\mu_{{\cal W}}^2)^2 p^2 
+ 36 \mu_{{\cal S}}^2 \mu_{{\cal W}}^2 \mu_{{\cal X}}^2 p^2 
+ 36 \mu_{{\cal S}}^2 \mu_{{\cal W}}^2 (p^2)^2 
+ 24 \mu_{{\cal T}}^2 (\mu_{{\cal W}}^2)^2 \mu_{{\cal X}}^2 
\right. \nonumber \\
&& \left. ~~~
+~ 24 \mu_{{\cal T}}^2 (\mu_{{\cal W}}^2)^2 p^2 
+ 48 \mu_{{\cal T}}^2 \mu_{{\cal W}}^2 \mu_{{\cal X}}^2 p^2 
+ 48 \mu_{{\cal T}}^2 \mu_{{\cal W}}^2 (p^2)^2 
- 18 (\mu_{{\cal W}}^2)^3 \mu_{{\cal X}}^2 - 18 (\mu_{{\cal W}}^2)^3 p^2 
\right. \nonumber \\
&& \left. ~~~
-~ 12 (\mu_{{\cal W}}^2)^2 \mu_{{\cal X}}^2 p^2 
- 12 (\mu_{{\cal W}}^2)^2 (p^2)^2 \right] \nonumber \\
&& \times 
\left[ [2 \mu_{{\cal Q}}^2 + 2 \mu_{{\cal R}}^2 - 2 \mu_{{\cal T}}^2 
+ 3 \mu_{{\cal W}}^2 - 2 p^2 ] [ \mu_{{\cal X}}^2 + p^2 ] + 6 \gamma^4 
\right]^{-1} 
\left[ \mu_{{\cal Q}}^2 - \mu_{{\cal T}}^2 + p^2 \right]^{-1} \nonumber \\
&& \times \left[ 2 \mu_{{\cal Q}}^2 + 3 \mu_{{\cal S}}^2 + 2 \mu_{{\cal T}}^2 
+ 3 \mu_{{\cal W}}^2 + 2 p^2 \right]^{-1} 
\left[ 2 \mu_{{\cal Q}}^2 + 3 \mu_{{\cal S}}^2 + 2 \mu_{{\cal T}}^2 
- 2 \mu_{{\cal W}}^2 + 2 p^2 \right]^{-1} \nonumber \\
{\cal L}_\xi &=& {\cal L}_\xi^L ~=~ {\cal L}_\rho ~=~ {\cal L}_\rho^L 
\nonumber \\ 
&=& 4 \left[ 2 \mu_{{\cal Q}}^2 \mu_{{\cal S}}^2 
+ 3 (\mu_{{\cal S}}^2)^2 + 2 \mu_{{\cal S}}^2 \mu_{{\cal T}}^2 
- \mu_{{\cal S}}^2 \mu_{{\cal W}}^2 + 2 \mu_{{\cal S}}^2 p^2 
- 2 (\mu_{{\cal W}}^2)^2 \right] \nonumber \\
&& \times \left[ 2 \mu_{{\cal Q}}^2 + 3 \mu_{{\cal S}}^2 + 2 \mu_{{\cal T}}^2 
+ 3 \mu_{{\cal W}}^2 + 2 p^2 \right]^{-1} 
\left[ 2 \mu_{{\cal Q}}^2 + 3 \mu_{{\cal S}}^2 + 2 \mu_{{\cal T}}^2 
- 2 \mu_{{\cal W}}^2 + 2 p^2 \right]^{-1} \nonumber \\
&& \times
\left[ 2 \mu_{{\cal Q}}^2 + \mu_{{\cal S}}^2 + 2 \mu_{{\cal T}}^2 
+ 2 \mu_{{\cal W}}^2 + 2 p^2 \right]^{-1} \nonumber \\
{\cal M}_\xi &=& {\cal M}_\xi^L ~=~ {\cal M}_\rho ~=~ {\cal M}_\rho^L 
\nonumber \\
&=& \frac{1}{2} \left[ 4 \mu_{{\cal P}}^2 (\mu_{{\cal Q}}^2)^2 
- 16 \mu_{{\cal P}}^2 \mu_{{\cal Q}}^2 \mu_{{\cal S}}^2 
+ 8 \mu_{{\cal P}}^2 \mu_{{\cal Q}}^2 \mu_{{\cal T}}^2 
- 6 \mu_{{\cal P}}^2 \mu_{{\cal Q}}^2 \mu_{{\cal W}}^2 
+ 8 \mu_{{\cal P}}^2 \mu_{{\cal Q}}^2 p^2 
- 33 \mu_{{\cal P}}^2 (\mu_{{\cal S}}^2)^2 
\right. \nonumber \\
&& \left. ~~~
-~ 16 \mu_{{\cal P}}^2 \mu_{{\cal S}}^2 \mu_{{\cal T}}^2 
+ 13 \mu_{{\cal P}}^2 \mu_{{\cal S}}^2 \mu_{{\cal W}}^2
- 16 \mu_{{\cal P}}^2 \mu_{{\cal S}}^2 p^2 
+ 4 \mu_{{\cal P}}^2 (\mu_{{\cal T}}^2)^2 
- 6 \mu_{{\cal P}}^2 \mu_{{\cal T}}^2 \mu_{{\cal W}}^2 
+ 8 \mu_{{\cal P}}^2 \mu_{{\cal T}}^2 p^2 
\right. \nonumber \\
&& \left. ~~~
+~ 14 \mu_{{\cal P}}^2 (\mu_{{\cal W}}^2)^2 
- 6 \mu_{{\cal P}}^2 \mu_{{\cal W}}^2 p^2 
+ 4 \mu_{{\cal P}}^2 (p^2)^2 
- 168 \mu_{{\cal Q}}^2 (\mu_{{\cal S}}^2)^2 
- 96 \mu_{{\cal Q}}^2 \mu_{{\cal S}}^2 \mu_{{\cal W}}^2 
- 252 (\mu_{{\cal S}}^2)^3 
\right. \nonumber \\
&& \left. ~~~
-~ 168 (\mu_{{\cal S}}^2)^2 \mu_{{\cal T}}^2 
- 12 (\mu_{{\cal S}}^2)^2 \mu_{{\cal W}}^2 
- 168 (\mu_{{\cal S}}^2)^2 p^2 
- 96 \mu_{{\cal S}}^2 \mu_{{\cal T}}^2 \mu_{{\cal W}}^2 
\right. \nonumber \\
&& \left. ~~~
+~ 144 \mu_{{\cal S}}^2 (\mu_{{\cal W}}^2)^2 
- 96 \mu_{{\cal S}}^2 \mu_{{\cal W}}^2 p^2 + 48 (\mu_{{\cal W}}^2)^3
\right] \nonumber \\
&& \times 
\left[ 2 \mu_{{\cal Q}}^2 + 15 \mu_{{\cal S}}^2 + 2 \mu_{{\cal T}}^2 
+ 6 \mu_{{\cal W}}^2 + 2 p^2 + 2 \mu_{{\cal P}}^2 \right]^{-1}
\left[ 2 \mu_{{\cal Q}}^2 + 3 \mu_{{\cal S}}^2 + 2 \mu_{{\cal T}}^2 
+ 3 \mu_{{\cal W}}^2 + 2 p^2 \right]^{-1} \nonumber \\
&& \times
\left[ 2 \mu_{{\cal Q}}^2 + 3 \mu_{{\cal S}}^2 + 2 \mu_{{\cal T}}^2 
- 2 \mu_{{\cal W}}^2 + 2 p^2 \right]^{-1} 
\left[ 2 \mu_{{\cal Q}}^2 + \mu_{{\cal S}}^2 + 2 \mu_{{\cal T}}^2 
+ 2 \mu_{{\cal W}}^2 + 2 p^2 \right]^{-1} \nonumber \\
{\cal N}_\xi &=& {\cal N}_\xi^L ~=~ {\cal N}_\rho ~=~ {\cal N}_\rho^L 
\nonumber \\
&=& \frac{1}{2} \left[ 16 (\mu_{{\cal Q}}^2)^2 \mu_{{\cal T}}^2 
+ 6 \mu_{{\cal Q}}^2 (\mu_{{\cal S}}^2)^2 
+ 56 \mu_{{\cal Q}}^2 \mu_{{\cal S}}^2 \mu_{{\cal T}}^2 
+ 24 \mu_{{\cal Q}}^2 \mu_{{\cal S}}^2 \mu_{{\cal W}}^2 
+ 32 \mu_{{\cal Q}}^2 (\mu_{{\cal T}}^2)^2 
+ 24 \mu_{{\cal Q}}^2 \mu_{{\cal T}}^2 \mu_{{\cal W}}^2 
\right. \nonumber \\
&& \left. ~~~
+~ 32 \mu_{{\cal Q}}^2 \mu_{{\cal T}}^2 p^2 + 9 (\mu_{{\cal S}}^2)^3 
+ 54 (\mu_{{\cal S}}^2)^2 \mu_{{\cal T}}^2 
+ 21 (\mu_{{\cal S}}^2)^2 \mu_{{\cal W}}^2 + 6 (\mu_{{\cal S}}^2)^2 p^2 
+ 56 \mu_{{\cal S}}^2 (\mu_{{\cal T}}^2)^2 
\right. \nonumber \\
&& \left. ~~~
+~ 40 \mu_{{\cal S}}^2 \mu_{{\cal T}}^2 \mu_{{\cal W}}^2 
+ 56 \mu_{{\cal S}}^2 \mu_{{\cal T}}^2 p^2 
+ 24 \mu_{{\cal S}}^2 \mu_{{\cal W}}^2 p^2 + 16 (\mu_{{\cal T}}^2)^3 
+ 24 (\mu_{{\cal T}}^2)^2 \mu_{{\cal W}}^2 + 32 (\mu_{{\cal T}}^2)^2 p^2 
\right. \nonumber \\
&& \left. ~~~
-~ 16 \mu_{{\cal T}}^2 (\mu_{{\cal W}}^2)^2 
+ 24 \mu_{{\cal T}}^2 \mu_{{\cal W}}^2 p^2 + 16 \mu_{{\cal T}}^2 (p^2)^2 
- 12 (\mu_{{\cal W}}^2)^3 \right] \nonumber \\
&& \times
\left[ 2 \mu_{{\cal Q}}^2 + 3 \mu_{{\cal S}}^2 + 2 \mu_{{\cal T}}^2 
+ 3 \mu_{{\cal W}}^2 + 2 p^2 \right]^{-1} 
\left[ 2 \mu_{{\cal Q}}^2 + 3 \mu_{{\cal S}}^2 + 2 \mu_{{\cal T}}^2 
- 2 \mu_{{\cal W}}^2 + 2 p^2 \right]^{-1} \nonumber \\
&& \times
\left[ 2 \mu_{{\cal Q}}^2 + \mu_{{\cal S}}^2 + 2 \mu_{{\cal T}}^2 
+ 2 \mu_{{\cal W}}^2 + 2 p^2 \right]^{-1}
\left[\mu_{{\cal Q}}^2 - \mu_{{\cal T}}^2 + p^2 \right]^{-1} \nonumber \\ 
{\cal A}^L &=& {\cal B}^L ~=~ {\cal C}^L ~=~ 0 \nonumber \\
{\cal K}_\xi^L &=& {\cal K}_\rho ~=~ {\cal K}_\rho^L \nonumber \\
&=& \frac{1}{3} \left[ 
8 (\mu_{{\cal Q}}^2)^2 \mu_{{\cal R}}^2 
+ 24 \mu_{{\cal Q}}^2 \mu_{{\cal R}}^2 \mu_{{\cal S}}^2 
+ 16 \mu_{{\cal Q}}^2 \mu_{{\cal R}}^2 \mu_{{\cal T}}^2 
- 8 \mu_{{\cal Q}}^2 \mu_{{\cal R}}^2 \mu_{{\cal W}}^2 
+ 16 \mu_{{\cal Q}}^2 \mu_{{\cal R}}^2 p^2 
+ 36 \mu_{{\cal Q}}^2 \mu_{{\cal S}}^2 \mu_{{\cal W}}^2 \right. \nonumber \\
&& \left. ~~~ 
+~ 48 \mu_{{\cal Q}}^2 \mu_{{\cal T}}^2 \mu_{{\cal W}}^2
- 12 \mu_{{\cal Q}}^2 (\mu_{{\cal W}}^2)^2 
+ 18 \mu_{{\cal R}}^2 (\mu_{{\cal S}}^2)^2 
+ 24 \mu_{{\cal R}}^2 \mu_{{\cal S}}^2 \mu_{{\cal T}}^2 
+ 6 \mu_{{\cal R}}^2 \mu_{{\cal S}}^2 \mu_{{\cal W}}^2 \right. \nonumber \\
&& \left. ~~~ 
+~ 24 \mu_{{\cal R}}^2 \mu_{{\cal S}}^2 p^2 
+ 8 \mu_{{\cal R}}^2 (\mu_{{\cal T}}^2)^2 
+ 16 \mu_{{\cal R}}^2 \mu_{{\cal T}}^2 \mu_{{\cal W}}^2 
+ 16 \mu_{{\cal R}}^2 \mu_{{\cal T}}^2 p^2 
- 12 \mu_{{\cal R}}^2 (\mu_{{\cal W}}^2)^2 \right. \nonumber \\
&& \left. ~~~ 
-~ 8 \mu_{{\cal R}}^2 \mu_{{\cal W}}^2 
+ 8 \mu_{{\cal R}}^2 (p^2)^2 
+ 27 (\mu_{{\cal S}}^2)^2 \mu_{{\cal W}}^2 
+ 36 \mu_{{\cal S}}^2 \mu_{{\cal T}}^2 \mu_{{\cal W}}^2 
+ 9 \mu_{{\cal S}}^2 (\mu_{{\cal W}}^2)^2 p^2 \right. \nonumber \\
&& \left. ~~~ 
+~ 36 \mu_{{\cal S}}^2 \mu_{{\cal W}}^2 
+ 24 \mu_{{\cal T}}^2 (\mu_{{\cal W}}^2)^2 
+ 48 \mu_{{\cal T}}^2 \mu_{{\cal W}}^2 p^2 
- 18 (\mu_{{\cal W}}^2)^3
- 12 (\mu_{{\cal W}}^2)^2 p^2 \right] \nonumber \\
&& \times 
\left[ 2 \mu_{{\cal Q}}^2 + 2 \mu_{{\cal R}}^2 - 2 \mu_{{\cal T}}^2 
+ 3 \mu_{{\cal W}}^2 + 2 p^2 \right]^{-1} 
\left[ \mu_{{\cal Q}}^2 - \mu_{{\cal T}}^2 + p^2 \right]^{-1} \nonumber \\
&& \times \left[ 2 \mu_{{\cal Q}}^2 + 3 \mu_{{\cal S}}^2 + 2 \mu_{{\cal T}}^2 
+ 3 \mu_{{\cal W}}^2 + 2 p^2 \right]^{-1} 
\left[ 2 \mu_{{\cal Q}}^2 + 3 \mu_{{\cal S}}^2 + 2 \mu_{{\cal T}}^2 
- 2 \mu_{{\cal W}}^2 + 2 p^2 \right]^{-1} 
\end{eqnarray}
Clearly there are no massless poles provided the combinations of the various
masses in each of the denominators do not accidentally sum to zero. Further,
the gluon propagator is transverse and freezes to a non-zero value in the
infrared when again there are no cancellations between the parameters.

\sect{${\cal W}$ and ${\cal S}$ channel propagators for $SU(N_c)$.}

As the explicit expressions for the propagators for each of the ${\cal W}$ and
${\cal S}$ channel cases are complicated for an arbitrary colour group we
present the expressions for $SU(N_c)$ only here. First, the ${\cal W}$ 
propagators are
\begin{eqnarray}
\left. \langle A^a_\mu(p) A^b_\nu(-p) \rangle_{{\cal W}} \right|_{SU(N_c)} 
&=& -~ \frac{\delta^{ab}[2p^2+N_c\mu_{{\cal W}}^2]}
{[2(p^2)^2+N_c\mu_{{\cal W}}^2p^2+2N_c\gamma^4]} P_{\mu\nu}(p) \nonumber \\
\left. \langle A^a_\mu(p) \xi^{bc}_\nu(-p) \rangle_{{\cal W}} \right|_{SU(N_c)} 
&=& \frac{2i f^{abc}\gamma^2}{[2(p^2)^2+N_c\mu_{{\cal W}}^2p^2+2N_c\gamma^4]} 
P_{\mu\nu}(p) \nonumber \\
\left. \langle A^a_\mu(p) \rho^{bc}_\nu(-p) \rangle_{{\cal W}} 
\right|_{SU(N_c)} &=& 0 \nonumber \\ 
\left. \langle \xi^{ab}_\mu(p) \xi^{cd}_\nu(-p) \rangle_{{\cal W}} 
\right|_{SU(N_c)} &=& -~ \frac{\delta^{ac}\delta^{bd} 
[4(p^2)^3+2N_c\mu_{{\cal W}}^2(p^2)^2-4\mu_{{\cal W}}^4p^2
-N_c\mu_{{\cal W}}^6]}
{2p^2[(p^2)^2-\mu_{{\cal W}}^4][2p^2+N_c\mu_{{\cal W}}^2]}\eta_{\mu\nu}
\nonumber \\
&& +~ \frac{2f^{ace} f^{bde} 
\mu_{{\cal W}}^2[3p^2+N_c\mu_{{\cal W}}^2]}
{3[(p^2)^2-(\mu_{{\cal W}}^2)^2][2p^2+N_c\mu_{{\cal W}}^2]}\eta_{\mu\nu}
\nonumber \\
&& -~ \frac{f^{abe} f^{cde}}{3p^2} 
\left[ 2[(N_c^2+6)\mu_{{\cal W}}^4p^2-3[2(p^2)^3-N_c\mu_{{\cal W}}^6]]\gamma^4
\right. \nonumber \\
&& \left. ~~~~~~~~~~~~~~~+~ 
\mu_{{\cal W}}^4p^2[N_cp^2+3\mu_{{\cal W}}^2][2p^2+N_c\mu_{{\cal W}}^2] \right]
\nonumber \\
&& ~~~~~~~~~~~~~~~ \times
\left[2(p^2)^2+N_c\mu_{{\cal W}}^2p^2+2N_c\gamma^4\right]^{-1}
\left[(p^2)^2-\mu_{{\cal W}}^4\right]^{-1} \nonumber \\
&& ~~~~~~~~~~~~~~~ \times 
\left[2p^2+N_c\mu_{{\cal W}}^2\right]^{-1} P_{\mu\nu}(p)
\nonumber \\
&& -~ \frac{f^{abe} f^{cde} 
\mu_{{\cal W}}^4[N_cp^2+3\mu_{{\cal W}}^2]}
{3p^2 [2p^2+N_c\mu_{{\cal W}}^2] [(p^2)^2-\mu_{{\cal W}}^4]} L_{\mu\nu}(p)
\nonumber \\
&& -~ \frac{2d_A^{abcd} \mu_{{\cal W}}^4}
{[2p^2+N_c\mu_{{\cal W}}^2] [(p^2)^2-\mu_{{\cal W}}^4]} \eta_{\mu\nu}
\nonumber \\
&& +~ \frac{\delta^{ab} \delta^{cd} N_c\mu_{{\cal W}}^6}
{[2p^2+N_c\mu_{{\cal W}}^2] [p^2+N_c\mu_{{\cal W}}^2]
[(p^2)^2-(\mu_{{\cal W}}^2)^2]} \eta_{\mu\nu} \nonumber \\
&& -~ \frac{\delta^{ad} \delta^{bc} N_c\mu_{{\cal W}}^6}
{2p^2[2p^2+N_c\mu_{{\cal W}}^2] [(p^2)^2-\mu_{{\cal W}}^4]} \eta_{\mu\nu} 
\nonumber \\
\left. \langle \xi^{ab}_\mu(p) \rho^{cd}_\nu(-p) \rangle_{{\cal W}} 
\right|_{SU(N_c)} &=& 0 \nonumber \\ 
\left. \langle \rho^{ab}_\mu(p) \rho^{cd}_\nu(-p) \rangle_{{\cal W}} 
\right|_{SU(N_c)} &=& 
\left. \langle \omega^{ab}_\mu(p) \bar{\omega}^{cd}_\nu(-p) \rangle_{{\cal W}}
\right|_{SU(N_c)} \nonumber \\
&=& -~ \frac{\delta^{ac}\delta^{bd} 
[4(p^2)^3+2N_c\mu_{{\cal W}}^2(p^2)^2-4\mu_{{\cal W}}^4p^2
-N_c\mu_{{\cal W}}^6]}
{2p^2[(p^2)^2-\mu_{{\cal W}}^4][2p^2+N_c\mu_{{\cal W}}^2]}\eta_{\mu\nu}
\nonumber \\
&& +~ \frac{2f^{ace} f^{bde} \mu_{{\cal W}}^4 [3p^2+N_c\mu_{{\cal W}}^2]}
{3[(p^2)^2-\mu_{{\cal W}}^4][2p^2+N_c\mu_{{\cal W}}^2]} \eta_{\mu\nu}
\nonumber \\
&& -~ \frac{f^{abe} f^{cde} \mu_{{\cal W}}^4 
[N_cp^2+3\mu_{{\cal W}}^2]}
{3p^2[(p^2)^2-(\mu_{{\cal W}}^2)^2][2p^2+N_c\mu_{{\cal W}}^2]} \eta_{\mu\nu}
\nonumber \\
&& -~ \frac{2d_A^{abcd} \mu_{{\cal W}}^4} 
{[(p^2)^2-(\mu_{{\cal W}}^2)^2][2p^2+N_c\mu_{{\cal W}}^2]} \eta_{\mu\nu}
\nonumber \\
&& +~ \frac{\delta^{ab} \delta^{cd} N_c\mu_{{\cal W}}^6} 
{[(p^2)^2-\mu_{{\cal W}}^4][2p^2+N_c\mu_{{\cal W}}^2]
[p^2+N_c\mu_{{\cal W}}^2]} \eta_{\mu\nu}
\nonumber \\
&& -~ \frac{\delta^{ad} \delta^{bc} N_c\mu_{{\cal W}}^6} 
{2p^2[(p^2)^2-\mu_{{\cal W}}^4][2p^2+N_c\mu_{{\cal W}}^2]} \eta_{\mu\nu} ~.
\label{propwsunc}
\end{eqnarray} 
Clearly the gluon propagator freezes to a non-zero value and the $\xi^{ab}_\mu$
and $\rho^{ab}_\mu$ propagators have massless poles in various colour channels.
More specifically the dominant part of each propagator as 
$p^2$~$\rightarrow$~$0$ is
\begin{eqnarray} 
\left. \langle \xi^{ab}_\mu(p) \xi^{cd}_\nu(-p) \rangle_{{\cal W}} 
\right|_{SU(N_c)} & \sim & -~ \frac{1}{2p^2} \left[ 
\delta^{ac} \delta^{bd} ~-~ \delta^{ad} \delta^{bc} ~-~ 
\frac{2}{N_c} f^{abe} f^{cde} \right] \eta_{\mu\nu} \nonumber \\ 
\left. \langle \rho^{ab}_\mu(p) \rho^{cd}_\nu(-p) \rangle_{{\cal W}} 
\right|_{SU(N_c)} & \sim & -~ \frac{1}{2p^2} \left[ 
\delta^{ac} \delta^{bd} ~-~ \delta^{ad} \delta^{bc} ~-~ 
\frac{2}{N_c} f^{abe} f^{cde} \right] \eta_{\mu\nu} ~. 
\end{eqnarray}
Interestingly the colour tensor structure of the dominant infrared behaviour of
each propagator is the same as that for the enhanced $\xi^{ab}_\mu$ propagator
in the pure Gribov-Zwanziger theory {\em after} the gap equation has been
implemented. Though the Lorentz structure differs. As the gluon propagator is 
essentially the same as that in the ${\cal R}$ channel case then the same 
features will arise when the one loop gap equation is computed for this case.
In other words the massless poles in the full set of ${\cal W}$ channel 
propagators will not enhance for similar reasons to those of the ${\cal R}$ 
case. Finally, just to be complete for this channel we note that the leading 
order behaviour of the adjoint colour projected $\xi^{ab}_\mu$ propagator as 
$p^2$~$\rightarrow$~$0$ is 
\begin{eqnarray}
\left. \langle f^{abp} \xi^{ab}_\mu(p) f^{cdq} \xi^{cd}_\nu(-p) 
\rangle_{{\cal W}} \right|_{SU(N_c)} & \sim & -~ \frac{2\delta^{pq}p^2}
{[2(p^2)^2+N_c\mu_{{\cal W}}^2p^2+2N_c\gamma^4]} P_{\mu\nu}(p) \nonumber \\
&& -~ \frac{2\delta^{pq}}{[2p^2+N_c\mu_{{\cal W}}^2]} L_{\mu\nu}(p) ~.
\end{eqnarray}
Clearly the transverse part of this particular correlator vanishes in the 
infrared but the longitudinal part freezes.

For the ${\cal S}$ channel the propagators are equally as involved since
\begin{eqnarray}
\left. \langle A^a_\mu(p) A^b_\nu(-p) \rangle_{{\cal S}} \right|_{SU(N_c)} 
&=& -~ \frac{\delta^{ab}p^2}
{[(p^2)^2+N_c\gamma^4]} P_{\mu\nu}(p) \nonumber \\
\left. \langle A^a_\mu(p) \xi^{bc}_\nu(-p) \rangle_{{\cal S}} \right|_{SU(N_c)} 
&=& \frac{i f^{abc}\gamma^2}{[(p^2)^2+N_c\gamma^4]} 
P_{\mu\nu}(p) \nonumber \\
\left. \langle A^a_\mu(p) \rho^{bc}_\nu(-p) \rangle_{{\cal S}} 
\right|_{SU(N_c)} &=& 0 \nonumber \\ 
\left. \langle \xi^{ab}_\mu(p) \xi^{cd}_\nu(-p) \rangle_{{\cal S}} 
\right|_{SU(N_c)} &=& \frac{ \delta^{ac} \delta^{bd}}{2p^2}
\left[ (N_c^2-36)N_c^2 \mu_{{\cal S}}^6 - 432(p^2)^3 - 72 (N_c^2+12) 
\mu_{{\cal S}}^2 (p^2)^2 \right. \nonumber \\
&& \left. ~~~~~~~~~-~ 6 (19N_c^2+72) \mu_{{\cal S}}^4 p^2 \right] \nonumber \\
&& ~~~~~~~~~ \times \left[ 6p^2 + N_c^2 \mu_{{\cal S}}^2 \right]^{-1}
\left[ 36(p^2)^2-(N_c^2-36)\mu_{{\cal S}}^4 \right]^{-1} \eta_{\mu\nu} 
\nonumber \\
&& -~ \frac{4f^{ace} f^{bde} N_c (N_c^2-9) \mu_{{\cal S}}^4}
{[6p^2+N_c^2\mu_{{\cal S}}^2] [36(p^2)^2-(N_c^2-36)\mu_{{\cal S}}^4]}
\eta_{\mu\nu} \nonumber \\
&& -~ \frac{f^{abe} f^{cde}}{p^2}
\left[ [ (N_c^2-36) \mu_{{\cal S}}^6 N_c^2 - 216 (p^2)^3 
- 36 (N_c^2+12) \mu_{{\cal S}}^2 (p^2)^2 \right. \nonumber \\
&& \left. ~~~~~~~~~~~~~~~~-~ 2 (N_c^2+18) (N_c^2+6) \mu_{{\cal S}}^4 p^2 ] 
\gamma^4 \right. \nonumber \\
&& \left. ~~~~~~~~~~~~~~~~-~ 2 (N_c^2-9) N_c \mu_{{\cal S}}^4 (p^2)^3 \right] 
\left[ (p^2)^2+N_c\gamma^4 \right]^{-1} \nonumber \\
&& ~~~~~~~~~~~~~~ \times \left[ 6p^2+N_c^2\mu_{{\cal S}}^2 \right]^{-1} 
\left[ 36(p^2)^2-(N_c^2-36)\mu_{{\cal S}}^4 \right]^{-1} P_{\mu\nu}(p) 
\nonumber \\
&& +~ \frac{2f^{abe} f^{cde} N_c (N_c^2-9) \mu_{{\cal S}}^4} 
{[6p^2+N_c^2\mu_{{\cal S}}^2] [36(p^2)^2-(N_c^2-36)\mu_{{\cal S}}^4]} 
L_{\mu\nu}(p) \nonumber \\
&& +~ \frac{12d_A^{abcd} [18p^2+(N_c^2+18)\mu_{{\cal S}}^2]\mu_{{\cal S}}^2}
{[6p^2+N_c^2\mu_{{\cal S}}^2] [36(p^2)^2-(N_c^2-36)\mu_{{\cal S}}^4]} 
\eta_{\mu\nu} \nonumber \\
&& -~ \frac{6\delta^{ab} \delta^{cd} N_c^2 
[126 p^2 + (5N_c^2+144) \mu_{{\cal S}}^2] \mu_{{\cal S}}^4}
{[6p^2+5N_c^2\mu_{{\cal S}}^2] [6p^2+N_c^2\mu_{{\cal S}}^2] 
[36(p^2)^2-(N_c^2-36)\mu_{{\cal S}}^4]} \eta_{\mu\nu} 
\nonumber \\
&& +~ \frac{\delta^{ad} \delta^{bc} N_c^2 
[18 p^2 - (N_c^2-36) \mu_{{\cal S}}^2] \mu_{{\cal S}}^4}
{2p^2 [6p^2+N_c^2\mu_{{\cal S}}^2] [36(p^2)^2-(N_c^2-36)\mu_{{\cal S}}^4]} 
\eta_{\mu\nu} \nonumber \\
\left. \langle \xi^{ab}_\mu(p) \rho^{cd}_\nu(-p) \rangle_{{\cal S}} 
\right|_{SU(N_c)} &=& 0 \nonumber \\ 
\left. \langle \rho^{ab}_\mu(p) \rho^{cd}_\nu(-p) \rangle_{{\cal S}} 
\right|_{SU(N_c)} &=& 
\left. \langle \omega^{ab}_\mu(p) \bar{\omega}^{cd}_\nu(-p) \rangle_{{\cal S}}
\right|_{SU(N_c)} \nonumber \\
&=& \frac{ \delta^{ac} \delta^{bd}}{2p^2}
\left[ (N_c^2-36)N_c^2 \mu_{{\cal S}}^6 - 432(p^2)^3 - 72 (N_c^2+12) 
\mu_{{\cal S}}^2 (p^2)^2 \right. \nonumber \\
&& \left. ~~~~~~~~~-~ 6 (19N_c^2+72) \mu_{{\cal S}}^4 p^2 \right] \nonumber \\
&& ~~~~~~~~~ \times \left[ 6p^2 + N_c^2 \mu_{{\cal S}}^2 \right]^{-1}
\left[ 36(p^2)^2-(N_c^2-36)\mu_{{\cal S}}^4 \right]^{-1} \eta_{\mu\nu} 
\nonumber \\
&& -~ \frac{4f^{ace} f^{bde} N_c (N_c^2-9) \mu_{{\cal S}}^4}
{[6p^2+N_c^2\mu_{{\cal S}}^2] [36(p^2)^2-(N_c^2-36)\mu_{{\cal S}}^4]}
\eta_{\mu\nu} \nonumber \\
&& +~ \frac{2f^{abe} f^{cde} N_c (N_c^2-9) \mu_{{\cal S}}^4}
{[ 6p^2+N_c^2\mu_{{\cal S}}^2 ] [ 36(p^2)^2-(N_c^2-36)\mu_{{\cal S}}^4 ]}
\eta_{\mu\nu} \nonumber \\
&& +~ \frac{12d_A^{abcd} [18p^2+(N_c^2+18)\mu_{{\cal S}}^2]\mu_{{\cal S}}^2}
{[6p^2+N_c^2\mu_{{\cal S}}^2] [36(p^2)^2-(N_c^2-36)\mu_{{\cal S}}^4]} 
\eta_{\mu\nu} \nonumber \\
&& -~ \frac{6\delta^{ab} \delta^{cd} N_c^2 
[126 p^2 + (5N_c^2+144) \mu_{{\cal S}}^2] \mu_{{\cal S}}^4}
{[6p^2+5N_c^2\mu_{{\cal S}}^2] [6p^2+N_c^2\mu_{{\cal S}}^2] 
[36(p^2)^2-(N_c^2-36)\mu_{{\cal S}}^4]} \eta_{\mu\nu} \nonumber \\
&& +~ \frac{\delta^{ad} \delta^{bc} N_c^2 
[18 p^2 - (N_c^2-36) \mu_{{\cal S}}^2] \mu_{{\cal S}}^4}
{2p^2 [6p^2+N_c^2\mu_{{\cal S}}^2] [36(p^2)^2-(N_c^2-36)\mu_{{\cal S}}^4]} 
\eta_{\mu\nu} ~. 
\label{propssunc}
\end{eqnarray} 
By contrast there is no gluon freezing but the massless poles are distributed
to the same colour channels as ${\cal W}$. However, the colour tensor structure
for the $\xi^{ab}_\mu$ propagator at leading order in the zero momentum limit 
is more akin to that observed for the enhanced $\xi^{ab}_\mu$ propagator in the
pure Gribov-Zwanziger case. Specifically as $p^2$~$\rightarrow$~$0$ we find 
that
\begin{eqnarray} 
\left. \langle \xi^{ab}_\mu(p) \xi^{cd}_\nu(-p) \rangle_{{\cal S}} 
\right|_{SU(N_c)} & \sim & -~ \frac{1}{2p^2} \left[ 
\delta^{ac} \delta^{bd} ~-~ \delta^{ad} \delta^{bc} \right] \eta_{\mu\nu} ~+~ 
\frac{1}{N_cp^2} f^{abe} f^{cde} P_{\mu\nu}(p) \nonumber \\ 
\left. \langle \rho^{ab}_\mu(p) \rho^{cd}_\nu(-p) \rangle_{{\cal S}} 
\right|_{SU(N_c)} & \sim & -~ \frac{1}{2p^2} \left[ 
\delta^{ac} \delta^{bd} ~-~ \delta^{ad} \delta^{bc} \right] \eta_{\mu\nu} 
\end{eqnarray}
where we include that for $\rho^{ab}_\mu$ for completeness. Whilst the same 
colour tensor structure emerges, the associated enhancement can only be 
determined when the one loop corrections are computed to all the $2$-point 
functions. If we examine the colour adjoint projection of the $\xi^{ab}_\mu$ 
propagator here then we find that at leading order in the zero momentum limit
\begin{equation}
\left. \langle f^{abp} \xi^{ab}_\mu(p) f^{cdq} \xi^{cd}_\nu(-p) 
\rangle_{{\cal S}} \right|_{SU(N_c)} 
~ \sim ~ -~ \frac{\delta^{pq}p^2}{[(p^2)^2+N_c\gamma^4]} P_{\mu\nu}(p) ~-~ 
\frac{\delta^{pq}}{p^2} L_{\mu\nu}(p) ~. 
\end{equation}
So the transverse part is finite and indeed vanishes as $p^2$~$\rightarrow$~$0$
whilst a massless pole is present in the longitudinal part. As discussed 
earlier we would expect that this will be enhanced when the loop corrections 
are included in the $2$-point functions since here the gluon propagator is
suppressed.

\end{document}